\documentclass[]{aa}
\usepackage{graphicx}
\usepackage{txfonts}
\usepackage{mathtools}
\usepackage{pdfpages}
\usepackage{natbib}
\usepackage{url}
\usepackage{placeins}
\usepackage{tikz}
\usepackage{wasysym}
\usepackage{gensymb}
\usepackage{ulem}
\usetikzlibrary{shapes,arrows}
\usepackage{geometry}
\usepackage{array}
\usepackage{colortbl}
\usepackage{xcolor}
\usepackage{pgf}
\usepackage{pgfmath}
\usepackage{verbatim}
\usepackage{amsmath}

\definecolor{myblue}{RGB}{0, 0, 140}
\usepackage{listings}
\lstnewenvironment{Ccode}
{\lstset{
  language=C,
  basicstyle=\ttfamily\scriptsize\color{myblue},
  frame=single,
  frameround=tttt,
  rulecolor=\color{myblue},
   framerule=0.5pt,
   tabsize=2
}} {}

\newcommand{\eq}{eq. }
\newcommand{\Eq}{Equation }

\newcommand{\sect}{Section }
\newcommand{\fig}{Figure }
\newcommand{\Fig}{Figure }

\newcommand{\applygradient}[2]{%
    \pgfmathsetmacro{\compA}{(1.0-((#1-#2)/(#2)))}%
    \edef\col{\noexpand\cellcolor[rgb]{1.0,\compA,\compA}}%
    \col #1%
}
\begin{document}

   \title{Optimized smoothing kernels for SPH}
   \author{Robert Wissing, Thomas Quinn, Ben Keller , James Wadsley, Sijing Shen 
          }
   \institute{
   Institute of Theoretical Astrophysics, University of Oslo, Postboks 1029, 0315 Oslo, Norway \\ \email{robertwi@astro.uio.no;}}
   \date{}
 
  \abstract
   { 
   We present a set of new smoothing kernels for smoothed particle hydrodynamics (SPH) that improve the convergence of the method without any additional computational cost. These kernels are generated through a linear-combination of other SPH kernels combined with an optimization strategy to minimize the error in the Gresho-Chan vortex test case. To facilitate the different choices in gradient operators for SPH in the literature, we perform this optimization for both geometric density average force SPH (GDSPH) and linear-corrected gradient SPH (ISPH). In addition to the Gresho-Chan vortex, we also perform simulations of the hydrostatic glass, Kelvin-Helmholtz instability and the Sod shocktube case. At low neighbour numbers ($<128$), there is a significant improvement across the different tests, with the greatest impact shown for GDSPH. Apart from the popular Wendland kernels we also explore other positive-definite kernels in this paper, which include the "missing" Wendland kernels, Wu kernels and Buhmann kernel. In addition, we also present a method for producing arbitrary non-biased initial conditions in SPH. This method uses the SPH momentum equation together with an artificial pressure combined with a global and local relaxation stage to minimize local and global errors.
  }

   \maketitle
%

\section{Introduction}
\label{sec:intro}
Smoothed particle hydrodynamics (SPH) is a Lagrangian particle method that has been applied to a wide range of topics within astrophysics. The traditional SPH method can be derived from the principle of least action and the Euler-Lagrange equations \citep{1988CoPhC..48...89M,2012JCoPh.231..759P}, resulting in a numerical scheme for hydrodynamics that spatially conserves linear momentum, angular momentum, entropy, and energy exactly, leaving the error in conservation mainly dependent on the time integration scheme \footnote{This is in regards with constant smoothing length or accounting for changes in smoothing length, $h$, through $\nabla h$ terms.}. Many of the drawbacks of the traditional SPH method, such as handling shocks, shear flow, convergence issues, and density gradients, have been greatly improved over the years. In addition, there are still numerous aspects of the method that remain unexplored, with the potential for even more improvements in the future. In this article we will focus on improving the convergence qualities of SPH. 
\\ \\
Since SPH is a particle-based method, the fluid is discretized by mass elements rather than volume elements as in grid-based codes. A smoothing kernel is used to interpolate the fluid quantities at any point in space. This includes the density, which determines the effective volume of the particles. While the sum of volume elements in a grid code remains constant the sum of particle volumes in an SPH code is highly dependent on the accuracy of the interpolation. This accuracy is mainly determined by the smoothing kernel and how well the particles are distributed within the kernel. An optimally distributed set of particles satisfies, for each particle $a$, the following constraints:
\begin{equation}
    Q_{0,a}=\sum_b \frac{m_b}{\rho_b} W_{ab}=1
\end{equation}
\begin{equation}
    Q_{1,a} = \sum_b \frac{m_b}{\rho_b} (\mathbf{r}_b - \mathbf{r}_a) W_{ab} = 0
\end{equation}
where the summation runs over neighboring particles $b$, with mass $m_b$, density $\rho_b$, and kernel weighting function $W_{ab}$. Here $Q_0$ represent the partition of unity and $Q_1$ indicates that there are no preferred direction or bias in the distribution of particles. The larger the deviation from these values, the worse the interpolation of fluid quantities. Related to this is the effective area between particles, which is analogous to the flux area between volume elements in grid-based codes. In a grid-based code the areas fully encloses the volume of the resolution element ($\int_s \hat{n} dS = 0$), where $S$ is the bounding surface of the resolution element and $\hat{n}$ is the unit normal vector pointing outward from the surface. However, in SPH this is not necessary true and will again depend on the particle distribution, such that we can have a non-cancelling of particle areas ($\int_s \hat{n} dS \neq 0$). As momentum flux is made to be symmetric in SPH (such that momentum is conserved), non-cancelling of particle areas leads to an automatic re-meshing mechanism in SPH. This causes particles to move to minimize this error and move toward a more optimal particle distribution
\footnote{When equal flux enters through each surface area between particles (like for constant pressure), a disparity in effective area on either side results in movement towards the side with the smaller effective area. This occurs because the sum of the effective areas between particles and their normals do not cancel out.}. This will occur even in the absence of hydrodynamic forces (constant pressure) and has thus been come to be known as the zeroth order error of SPH. Particles that are optimally distributed fulfill the cancellation of areas, and also the following conditions:
\begin{equation}
    E_{0,a}^i=2\sum_b \frac{m_b}{\rho_b} \overline{\nabla W_{ab}}=0 
    \label{eq:E0}
\end{equation}
\begin{equation}
    E_{1,a}^{ij}=2\sum_b \frac{m_b}{\rho_b} (\mathbf{r}_b - \mathbf{r}_a)^i \overline{\nabla_a^j W_{ab}}=\delta^{ij}
    \label{eq:E1}
\end{equation}
Here $E_0$ is the zeroth order error \footnote{gradient of the smoothing kernel give the effective area between two particles} and $E_1$ is the linear gradient error. The indices $i,j \in \{x,y,z\}$ denote Cartesian vector components and $\delta^{ij}$ the Kronecker delta. Here we have used the GDSPH gradient operator \citep{2017MNRAS.471.2357W}, which we will continue to use for the rest of this paper. The momentum equation for SPH is derived from the Euler equations and given by:
\begin{equation}
    \frac{dv}{dt}=\frac{-\nabla P}{\rho} \rightarrow \frac{dv_a}{dt}=-\sum_{b} m_b \left( \frac{P_a+P_b}{\rho_a\rho_b} \right) \overline{\nabla W_{ab}}
\end{equation}
To get the error terms of this function we can Taylor expand $P_b$ around $r_a$: 
\begin{equation}
\begin{split}
 \frac{dv_i}{dt} &= \frac{2P_a}{\rho_a}\sum_b \frac{m_b}{\rho_b} \overline{\nabla_a^i W_{ab}} \\ 
 & +\frac{2\nabla_{i} P_a}{\rho_a}\sum_b \frac{m_b}{\rho_b} (r_{b}-r_a)^{j} \overline{\nabla_a^j W_{ab}} + O(h^2).
\end{split}
\end{equation}
We can see that our momentum equation is second order if the particle fulfills the $E_0$ and $E_1$ conditions. Most of the time this is not the case, but we can develop optimizations to try and minimize these kinds of errors. These optimization often involve different smoothing kernels, number of neighbours, corrections to the gradients, different gradient operators, or different choices of effective particle volume. 
\\ \\
Linear-exact gradient corrections have become popular in recent times, due to the increased accuracy they offer in subsonic flows and in shearing flows. Here, a matrix inversion of the $E_1$ condition is performed and then applied to the gradient of the kernel to ensure that the $E_1$ condition is fulfilled. There are a few variations to this correction. The most straightforward is to simply invert $E_1$ in \eq~\ref{eq:E1} \citep{1999CMAME.180...97B,2004MNRAS.348..123P}. In the CRKSPH formulation \citep{2017JCoPh.332..160F,2024arXiv241119228R}, the kernel and its derivatives are reproduced to enforce both the zeroth order and the linear order; however, when deriving the pair-wise forces to uphold linear momentum conservation, the zeroth order correction is effectively lost, and only the linear correction remains. There might be benefits in these reproducing kernels over other linear correction methods, but that remains to be seen. Another way of constructing linear-exact gradients for SPH was devised by \cite{2012A&A...538A...9G}. In their proposal, gradients are calculated from an integral expression, so that there is no need to explicitly calculate the analytic derivative of the kernel function. Similarly to \cite{1999CMAME.180...97B,2004MNRAS.348..123P}, a matrix inversion is then performed to linearly correct for the gradients. This method is known as the integral smoothed particle hydrodynamics (ISPH) scheme \citep{2012A&A...538A...9G}. Integral-based estimates have been applied readily in the past for second derivatives as they are much less noisy than analytical second derivatives of the smoothing kernel \citep{1985PASA....6..207B,2005RPPh...68.1703M}. ISPH have shown great advantages over regular SPH in subsonic flows \citep{2012A&A...538A...9G,2015MNRAS.448.3628R,2016ApJ...831..103V}. The disadvantages of these linear corrections include the additional computational cost and the discrepancy in errors between the density estimate and velocity gradients, potentially leading to entropy errors. Global angular momentum is also no longer fully conserved in these formulation as the pair-wise force is not always radially aligned between particles. This becomes a resolution dependent error, and has been stated to remain relatively small \citep{2017JCoPh.332..160F}; however, this error is dependent on both resolution and the given particle distribution. On the other hand, this is offset by more precise local angular momentum transport due to more accurate gradients.
\\ \\
The interpolation kernel is the foundation of the SPH method and is ever present in determining the accuracy of a simulation, even when applying the optimizations mentioned above. For a long time B-splines were the most popular smoothing kernel to use, due to their compact support, good interpolation and flexibility in polynomial degree. However, an issue with these kernels has always been that they are susceptible to the pairing-instability. The pairing-instability occurs when the number of neighbors exceeds a certain critical value, causing particles to clump together and consequently reduce the resolution of the simulation \citep{1992MNRAS.257...11T,1996PASA...13...97M,2004ApJS..153..447B,2012JCoPh.231..759P,2012MNRAS.425.1068D}. This will depend on the kernel, but for the cubic kernel it lies in the region of $N_{smooth}=50$. Increasing the number of neighbors leads to more accurate low order errors ($E_0,E_1,Q_0,Q_1$) and less particle noise. This is particularly desirable when modeling subsonic flows where smooth velocity fields are essential \footnote{The drawbacks of using a high number of neighbors is the additional computational cost and the increase in smoothing length ($h$), which can lead to excessive smoothing of sharp features in the flow.}. This is why Wendland kernels have become very popular in recent times, as they are stable against the pairing instability at all neighbour numbers. It was shown in \cite{2012MNRAS.425.1068D} that a positive Fourier transform is a necessary condition for stability against the pairing instability, though it is still unclear why this triggers the pairing instability. It has been hypothesized that this is a consequence of the particles trying to re-order themselves to minimize the total internal energy ($U$) given a fixed entropy. While the uniform particle distribution always represents a local minimum, if a paired particle distribution results in a lower internal energy, then the uniform particle distribution is only meta-stable. Cubic spline kernels are hypothesized to exhibit this property as it has an oscillating over/under estimation of density given a uniform particle distribution for varying neighbor numbers. While the Wendland kernel and other kernels with positive definite Fourier transform seem to exhibit an overestimation of the density, which continuously decrease as number of neighbors increase. \cite{2012MNRAS.425.1068D} hypothesizes that pairing occurs when the following condition is fulfilled (pairing occurs if $\rho(N_{smooth}/f)<\rho(N_{smooth})$ for some $1<f\leq2$). The overestimation of the density by the Wendland kernel can be especially high when using low neighbour numbers. One can however, correct for the kernel bias by adjusting the self-interaction term ($W(0,h)$), effectively removing this bias from the density estimate while leaving the gradients unaffected. While the Wendland kernels have been popularized, there are many other interpolation kernels that meet the criteria of having a positive Fourier transform, Gaussian shape and compact support. These include the Wu family of kernels \citep{articleWU}, the missing Wendland kernels \citep{articleMISSWEND} and the Buhmann family of kernels \citep{articleBUH}. Another family of kernels is the Sinc kernels $W(q) = \left( \sin(\frac{\pi}{2} q) / (\frac{\pi}{2} q) \right)^n$ \citep{2008JCoPh.227.8523C,2014A&A...570A..14G}, which is directly linked to the Dirac-$\delta$ function. Sinc kernels, like the B-spline family of kernels, can be subject to the pairing instability. However, the critical neighbour number for the instability can be pushed higher by increasing the sharpness of the kernel, which is controlled by the kernel exponent $n$. A consequence of increasing the sharpness is that it usually gives worse interpolation at lower neighbour numbers. This is likely related to the larger difference in weight between the inner and outer regions produced by sharper kernels. Recently, a linear combination of two Sinc kernels were proposed by \cite{2024MNRAS.528.3782C}, to generate a kernel that is more resistant to the pairing instability, while retaining good interpolation properties over a wide range of neighbours numbers. In this work, we aim to design smoothing kernels specifically optimized for SPH, rather than using interpolation kernels developed for other purposes. Our approach involves optimizing a linear combination of kernels, similar to the method in \cite{2024MNRAS.528.3782C}, to produce the most accurate kernel for a given neighbor number. A big challenge during this work was to determine the most natural initial conditions to minimize any kind of bias.
\\ \\
Lattice initial conditions are often used to provide an initially optimal distribution of particles. Here, particles are put in a cubic or closed package lattice. This might at first seem a nice solution as the $E_0$ and $E_1$ conditions are fulfilled. But there are many disadvantages to actually using a lattice for the initial condition. First, the lattice structure is easily broken by shocks and shear flows, which will generate aggressive noise as the particles move off the lattice structure, and this will cascade through the simulation volume. Second, particle lattice distributions can introduce directional bias and simulation artifacts. For instance, a shock wave propagating along a preferred direction of the lattice can disproportionately gather particles in that direction, leading to undesirable oscillations. Third, the two issues above are exacerbated when dealing with density gradients in the initial conditions. Lattice particle distributions can give both an artificial positive result for SPH, as it provides very high accuracy as long as the lattice holds up, but also an artificial negative result, as significant noise is generated when the lattice breaks down. A much more natural distribution for SPH is the glass distribution, which SPH will always strive to re-mesh towards as particles are disturbed by shocks, shear flows, etc. Glass distributions are often obtained by letting the particles relax, beginning with a lattice or random distribution, adding some random velocities and using a velocity damping term to eventually approach a steady state. This can be done with the influence of gravity to generate collapsed density structures or without to generate uniform distributions. However, this method has both limitations and downsides. First, one cannot generate density contrasts for setups that are purely hydrodynamic\footnote{One can take two glass boxes with different densities and put them next to each other; however, here the interface is not in a relaxed distribution. And if you want to smooth the density contrast this also becomes problematic.} (the Kelvin-Helmholtz setup for example); second, glass generation in this way can be costly as oscillations can take a long time to damp; and third, damping the velocities too quickly can lead to inaccurate distributions. In this paper, we have instead developed a glass generator that relies on the SPH method to find a relaxed glass distribution for arbitrary density gradients.
\\ \\
The use of lattice initial conditions is especially problematic when it comes to assessing the quality of a smoothing kernel. As the results do not reflect the natural distribution of particles and give a positive bias towards smoother kernels, which usually have larger $E_0$ errors in glass distributions (given the same neighbour count). In \cite{2012MNRAS.425.1068D}, the correction for the kernel bias was made based on the bias when interpolating a closed-packed lattice distribution. The bias was also given by a simple power law. In this paper we calculate a new correction to the kernel bias for all our kernels, which is based on the generated glass that each kernel itself produces. We also give a piecewise function to capture the kernel bias more accurately over a wider-range of neighbour numbers.
\\ \\
In \sect \ref{sec:kernelprop}, we first present all the smoothing kernels used in this paper. In \sect \ref{sec:Method}, we go through the methods used in this paper, this includes the SPH methods, the IC generator, adjustment of density bias and optimization strategy. In \sect \ref{sec:results} we present the results from the tests. And finally in \sect \ref{sec:discussion} we discuss our results.
\section{Kernel properties}
\label{sec:kernelprop}
All the kernels that we go through in this section go to 0 at 
\begin{equation}
    q=(r/2h)>1,
\end{equation} even if not stated explicitly in the equations. All kernels are normalized with:
\begin{equation}
    \sigma = \frac{1}{\int_0^{1}\int_0^{\pi}\int_0^{2\pi} W(r) r^2 sin(\theta) dr d\theta d\psi }
\end{equation}
A useful tool for analyzing kernels is their Fourier transform. Despite having similar shapes, kernels can exhibit quite different Fourier transforms. As previously mentioned, negative values in the Fourier transform are responsible for the pairing instability. The lower the wave numbers at which these negative values appear, the lower the critical number of neighbors required to trigger this instability. The Fourier transform is defined as:
\begin{equation}
    F_3[W(r)](k)=4\pi k^{-1} \int_0^{\infty} sin(kr)W(r)rdr
\end{equation}
where $W(r)$ is the smoothing kernel and $k$ is the wave number. All the kernels that we mention in this section are positive-definite, which means that they have as strictly positive Fourier transform. Positive definite kernels all produce a general over-bias in their density estimation, which become more prominent at low neighbor numbers, as the bias is dominated by the self contribution. The neighbour number at which this bias becomes negligible depends on the smoothing kernel (higher order/more peaked kernels give higher bias). This self contribution can be corrected for as described in \sect~\ref{sec:densitybias}.
\subsection{Generalized Wendland kernel}
\label{sec:genwendkern}
The Wendland kernels is a family of compactly supported radial functions, generated by a dimension walk (integration) of a truncated power function \citep{Wendland1995}.
$$\psi_l(r)=(1-r)^l_+$$
$$l(k) = \left\lfloor 1.5 + k \right\rfloor + 1$$
\begin{equation}
   \phi_k(r)=\mathcal{I}^k\psi_{l} 
\end{equation}
Here $k$ is the smoothness parameter, $l(k)$ is the order of the kernel and $\lfloor x \rfloor$ is the floor operator. We have set $l(k)$ so that the generated function is strictly positive definite and radial on $R^3$ (following $D \leq 2l-1$ for $D$ dimensions). The Wendland kernels produce the minimal polynomial degree for a given space dimension and smoothness.
While initially only determined for positive integer smoothness parameters, this has been generalized for non-integer smoothness parameters in the generalized Wendland functions \citep{chernih_closed_2014}:
\begin{multline}
\phi_k(r) = \frac{\Gamma[l(k) + 1]}{2^{l(k) + k} \Gamma[l(k) + k + 1]} \left(1 - r^2\right)^{l(k) + k} r^{-l(k)} \\
\times \, {}_2F_1\left(\frac{l(k)}{2}, k + \frac{l(k) + 1}{2}, l(k) + k + 1, 1 - \frac{1}{r^2}\right)
\end{multline}
Here $\Gamma$ is the gamma function and ${}_2F_1$ is the hypergeometric function. The family of Wendland kernel functions for 3D is then defined as:
\begin{equation}
\label{eq:smoothkern4}
    W(r)=\sigma \phi_k(r)
\end{equation}
Positive integers of $k$ represent the classic Wendland kernels, where the resulting kernels have polynomial structure. The Wendland kernels with positive integers used in this paper ($k=0,1,2,3$) are:
\begin{equation}
\label{eq:WC0}
    W_{C0}(q)=\sigma(1-q)^2
\end{equation}
\begin{equation}
\label{eq:WC2}
    W_{C2}(q)=\sigma\left((1-q)^4(1+4(1-q))\right)
\end{equation}
\begin{equation}
\label{eq:WC4}
    W_{C4}(q)=\sigma\left((1-q)^6(1+6(1-q)+\frac{35}{3}(1-q)^2)\right)
\end{equation}
\begin{equation}
\label{eq:WC6}
    W_{C6}(q)=\sigma\left((1-q)^8(1+8(1-q)+25(1-q)^2+32(1-q)^3)\right)
\end{equation}
Positive half-integers are known as the missing Wendland kernels \citep{articleMISSWEND}. They are compactly supported and polynomial but with additional logarithmic and square-root terms ($k=0.5$):
\begin{multline}
W_{CM_{05}} = \sigma \left( \sqrt{1 - q^2} + 6.5 \, q^2 \sqrt{1 - q^2} \right. \\
\left. + 1.5 \, q^2 (4 + q^2) \left( \log{q} - \log{(1 + \sqrt{1 - q^2})} \right) \right)
\end{multline}
Positive non-integer kernels can also be generated; these involve a more complicated form ($k=0.4$):
\begin{equation}
W_{CM_{04}} = \sigma\frac{0.375 (1 - q^2)^{2.4} \, {}_2F_1\left(1, 1.9, 3.4, 1 - \frac{1}{q^2}\right)}{q^2}
\end{equation}
To avoid handling hyperbolic functions within our code we approximate this function with a polynomial fit. Higher-order kernels have smoother derivatives, which in combination with higher neighbor numbers act to decrease the sensitivity to particle disorder. The trade-off of using a higher-order kernel is that it becomes less accurate than its lower-order counterpart at low neighbor numbers \citep{1992ARA&A..30..543M,2015MNRAS.448.3628R}. We will refer to the Wendland kernels as C2, C4, C6, CM05, CM04 throughout this paper.
\subsection{Wu kernel}
\label{sec:wukernel}
The Wu approach starts with the function \citep{articleWU}:
\begin{equation}
    \phi=(1-r^2)_+^{l}.
\end{equation}
Another function is then generated by convolution:
\begin{equation}
    \phi_\ell(r) =\int_{-1}^1 (1 - t^2)^\ell (1 - (2r - t)^2)^\ell \, dt.
\end{equation}
Using this function the Wu family of kernels is generated by dimension walk (derivative):
\begin{equation}
    \phi_{l,k}=\mathcal{D}^k\phi_{l}.
\end{equation}
The Wu kernel ($\phi_{l,k}$) is strictly positive definite and radial in $R^D$ for $D\leq 2k+1$, with polynomial of degree $4l-2k+1$ and smoothness of $C^{2(l-k)}$, which indicates how many times the kernel is continuously differentiable ($2(l-k)$ times). The polynomial degree is higher for the Wu kernel than the Wendland kernel at a certain smoothness. In this paper we have chosen to look at the $\phi_{1,2}$ and $\phi_{1,3}$:
\begin{equation}
W_{WU2}=(1-q)^4_+ (4 + 16 q + 12 q^2 + 3 q^3)
\end{equation}
\begin{equation}
W_{WU4}=(1-q)^6_+ (6 + 36q + 82q^2 + 72q^3 + 30q^4 + 5q^5)
\end{equation}
For a prescribed smoothness the polynomial degree of Wendland kernels are lower than that of Wu kernels. An interesting thing about Wu kernels is that their Fourier transform is positive, quickly decaying and goes to zero in periodic intervals. We will refer to the Wu kernels as WU2 and WU4 throughout this paper.
\subsection{Buhmann kernel}
\label{sec:buhkernel}
Buhmann’s approach \citep{articleBUH} involves integrating a positive function \( f(t) = t^\alpha (1 - t^\delta)_+^\rho \) against a strictly positive definite kernel \( K(t, r) = \left(1 - \frac{r^2}{t}\right)_+^\lambda \). Buhmann's family of kernels thus take the general form:
\begin{equation}   \phi(r)=\int^{\infty}_{0}(1-r^2/t)_+^{\lambda} t^{\alpha} (1-t^{\delta})^{\rho}_+ dt
\end{equation}
Here $0<\delta\leq0.5$, $\rho>0.5$ and the kernel is strictly positive and radial in $R^D \leq 3$ for $\lambda \geq 0$ and $-1<\alpha<\frac{\lambda-1}{2}$.
Buhmann's family of kernels consists of a polynomial and an additional logarithmic term. The Buhmann kernel that we include in this paper is ( \(\alpha = \delta = \frac{1}{2}\) \(\rho = 1\) \(\lambda = 2\) from \cite{2003rbf..book.....B}):
\begin{equation}
W_{Buh}(r) = \sigma \left( 12q^4 \log q - 21q^4 + 32q^3 - 12q^2 + 1 \right).
\end{equation}
We will refer to this kernel as BUH throughout this paper.
\section{Method}
\label{sec:Method}
\subsection{SPH methods}
\label{sec:SPHmethods}
We will perform the optimization using both the GDSPH and ISPH methods. We do this because the ``best'' kernel may differ between the two methods due to the removal of the linear error term. The density of both GDSPH and ISPH is calculated using the standard density estimate:
\begin{equation}
    \rho_a = \sum_b m_b W_{ab}(h_a).
    \label{eq:densest}
\end{equation}
\subsubsection{GDSPH gradient}
\label{sec:gdsph}
We implement the GDSPH gradient operators of Gasoline2 \citep{2017MNRAS.471.2357W} in this paper. This variant of SPH is identical to regular SPH when the density is uniform, but improves gradient accuracy in cases involving sharp density gradients.
\begin{equation}
    \frac{d \mathbf{v}_a}{dt} = - \sum_b m_b \left[ \frac{f_a P_a+f_b P_b}{\rho_a\rho_b} \overline{\nabla_a W_{ab}} \right],
    \label{eq:dvdtgd}
\end{equation}
\begin{equation}
    \frac{d u_a}{dt} = \frac{f_a P_a}{\rho_a} \sum_b \frac{m_b}{\rho_b} \mathbf{v}_{ab} \cdot \overline{\nabla_a W_{ab}}.
    \label{eq:dudtgd}
\end{equation}
Here $f_a$ are entropy corrections terms to remove the linear entropy error, caused by a mismatch between GDSPH operators and the density estimate.
\subsubsection{Integral based gradient}
\label{sec:isph}
In ISPH \citep{2012A&A...538A...9G}, the gradients are calculated from an integral expression, so that there is no need to explicitly calculate the analytic derivative of the kernel. The kernel gradients are also linear-corrected by the use of a matrix inversion of the $E_1$ error. Effectively we replace the kernel gradients of \Eq \ref{eq:dvdtgd} and \ref{eq:dudtgd} with ($\overline{\nabla_a W_{ab}} \rightarrow \overline{G_{ab}^k}$):
\begin{equation}
(G_{a,j})^k = \sum_{d=1}^{3} C^{kd}_a r_{ba}^d W_{ab}(h_j),
\end{equation}
\begin{equation}
(G_{b,j})^k = \sum_{d=1}^{3} C^{kd}_b r_{ba}^d W_{ab}(h_j),
\end{equation}
where $j$ is either $a$ or $b$.
\begin{equation}
    \overline{G_{a}^k}=0.5(G_{a,a}^k+G_{a,b}^k)
\end{equation}
\begin{equation}
    \overline{G_{b}^k}=0.5(G_{b,a}^k+G_{b,b}^k)
\end{equation}
The linear correction matrix, $C$ is calculated with:
\begin{equation}
C^{ki}_a = \left( \sum_b \frac{m_b}{\rho_b} r_{ba}^k r_{ba}^i \overline{\nabla_a W_{ab}} \right)^{-1}.
\label{eq:Cisph}
\end{equation}
This form is slightly different than the one proposed by previous authors\citep{2012A&A...538A...9G,2020MNRAS.498.4230R}. This is because it involves the symmetric kernel $\overline{\nabla_a W_{ab}}$ within the linear correction matrix, instead of the one-sided kernel gradient $\nabla_a W_{ab}(h_a)$. The momentum and internal energy equations become:
\begin{equation}
    \frac{d \mathbf{v}_a}{dt} = - \sum_b m_b \left[ \frac{ P_a \overline{G_{a}^k}+ P_b \overline{G_{b}^k}}{\rho_a\rho_b} \right],
    \label{eq:dvdtisph}
\end{equation}
\begin{equation}
    \frac{d u_a}{dt} = \frac{P_a}{\rho_a} \sum_b \frac{m_b}{\rho_b} \mathbf{v}_{ab} \cdot \overline{G_{a}^k} .
    \label{eq:dudtisph}
\end{equation}
This matrix inversion requires us to do an extra neighbour loop and requires us to save six additional quantities per particle (symmetric matrix). This linear correction can introduce a global angular momentum error as pairwise torques can be generated between particles. This is because the forces between particles are no longer ensured to be aligned along the line connecting particles. This is in general true for any kernel correction of order greater than zero. This effectively results in a resolution/kernel dependent error in global angular momentum conservation.
\subsection{Initial condition - glass generator}
\label{sec:ICgen}
\begin{figure}
    \centering
    \includegraphics[width=\linewidth]{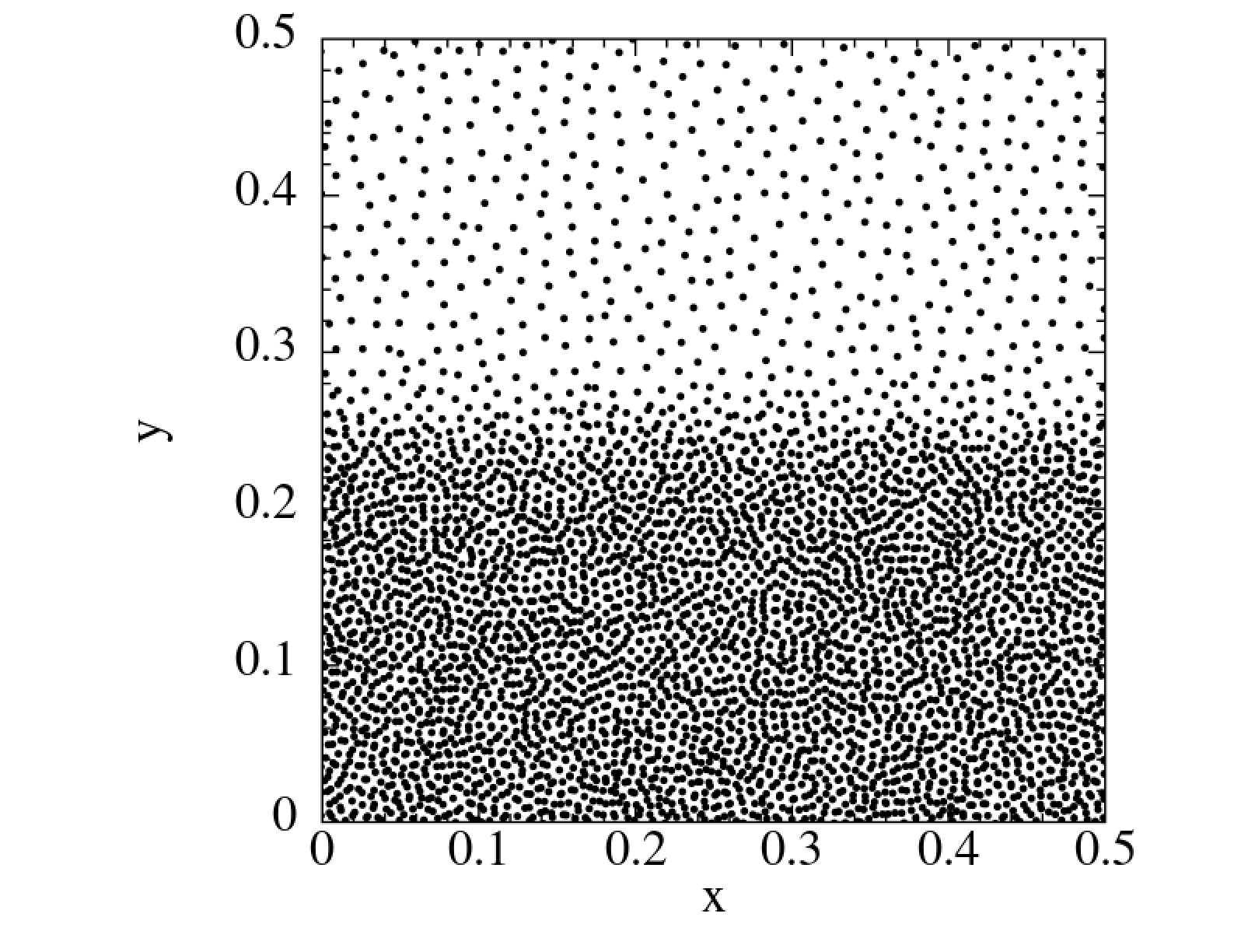}
    \includegraphics[width=\linewidth]{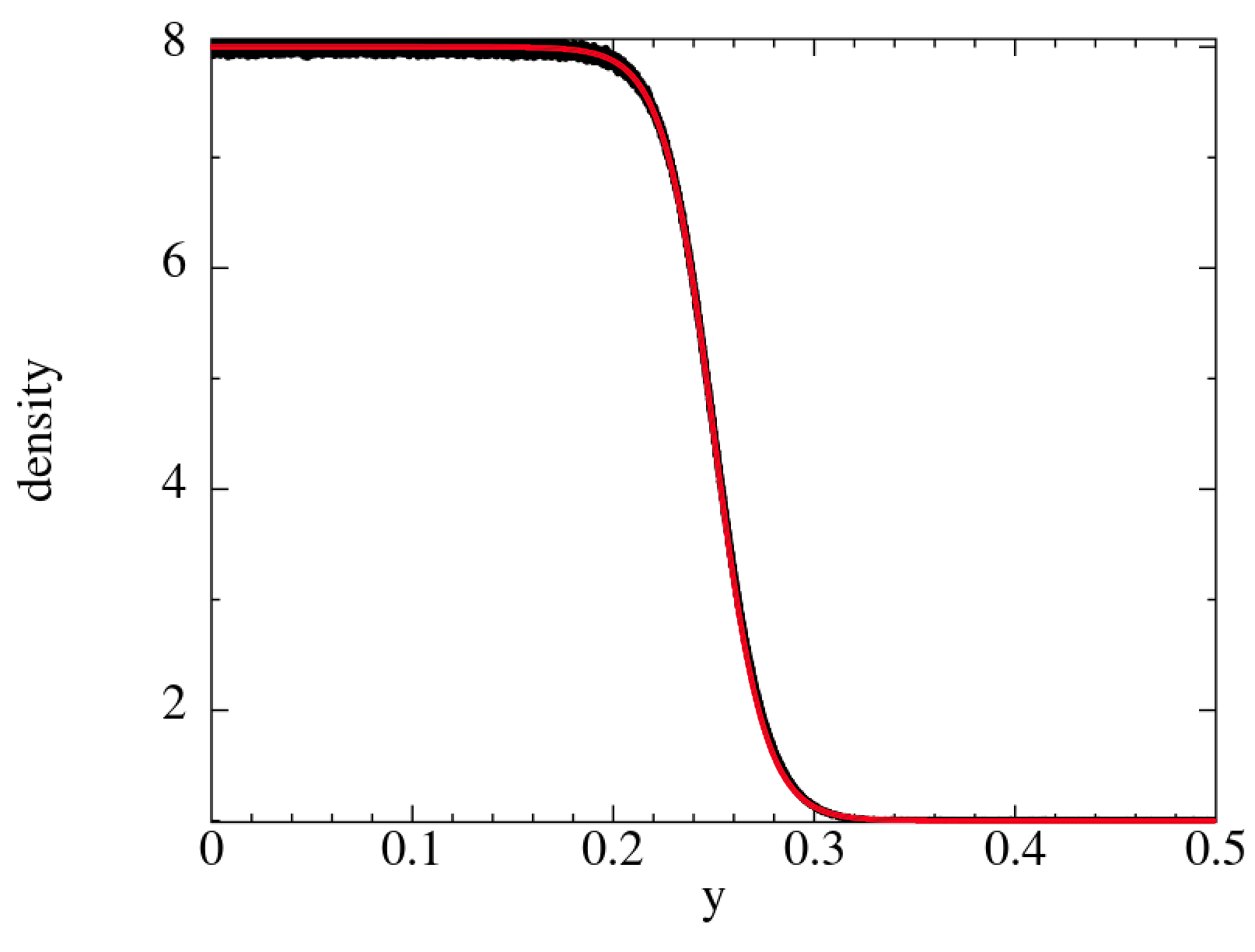}
    \caption{Relaxed glass IC generated by the IC generator outlined in \sect \ref{sec:ICgen}. Here we model the IC of the Kelvin Helmholtz instability (\sect \ref{sec:reskelvin}), with a   density contrast of $\Delta \rho= 8$. The top panel illustrates the particle distribution within a thin slice of thickness $\Delta z = 0.015$. The bottom panel show the density profile in the y direction. The red line depicts the analytical density profile and the black dots the density of the particles. The glass generated density profile is in excellent agreement with the analytical solution.  }
    \label{fig:ICGEN}
\end{figure}
As discussed in the introduction, producing natural and non-biased initial conditions are important for SPH. Of particular difficulty is the generation of non-uniform density distributions. Lattice initial conditions can be used in non-uniform density distributions, by either varying particle mass \citep{Lombardi1999} or stretching a uniform lattice \citep{2009NewAR..53...78R, 2018PASA...35...31P}. The main issue with these methods are the use of the lattice initial condition, which as discussed are artificial, biased and experience excessive noise when the lattice breaks. Varying particle masses are undesirable as well, due to particles having more difficulty in balancing the particle distribution and generally leading to larger gradient errors and noise. In \cite{2015PASA...32...48D} a more general initial condition generator is proposed, using the concept of weighted Voronoi tessellation. This works well in generating non-uniform density distributions. The main issue with this approach is that the generated distribution does not necessarily correspond to the same natural particle distribution that a given SPH method would produce, as it does not take into account the smoothing kernel or the gradient operators. This means that it will also represent a slightly biased initial condition. In \cite{2020MNRAS.498.4230R}, a more SPH-like version of this method is applied, where the SPH momentum equation is used, together with an artificial pressure based on the current density error. This is an excellent approach for generating non-biased initial conditions for SPH, as it ensures a near noise free distribution at the start of your simulation (as one is using the same momentum equation to evolve the system). We propose a similar method in this paper, but add some important alterations to ensure that both local and global errors are minimized. In addition, this method will relax to a given density profile, no matter what your starting initial condition is (given that the mass of the system is correct). Similar to \cite{2020MNRAS.498.4230R}, we use an iterative method where during each iteration, $i$, we update the position directly for each particle:
\begin{equation}
    \mathbf{r}_{a,i+1} = \mathbf{r}_{a,i} + \Delta \mathbf{r}_{a,i}
\end{equation}
The displacement of each particle is determined by the following equations:
\begin{equation}
    \Delta r_{a,i} = K_{global,i}\frac{dv_i/dt}{f_{max,i}} h_i
\end{equation}
\begin{equation}
    f_{max,i+1}=\max\left(\left\lvert\frac{dv}{dt}\right\lvert\right).
\end{equation}
$K_{global,i}$ will determine the largest multiple of the smoothing length that a particle can be displaced. The variables $\frac{dv}{dt}$ and $\rho_i$ are calculated with the chosen density estimate and gradient operator (in this paper GDSPH (\eq \ref{eq:dvdtgd}) and ISPH (\eq \ref{eq:dvdtisph})).  Hence the produced glass will represent the relaxed distribution of the chosen gradient operator, smoothing kernel and neighbour number. The pressure in the momentum equation is replaced by an artificial pressure, given by:
\begin{equation}P_a=\left(\frac{\rho_{tar}}{\rho_a}\right)^n.
\end{equation}
Here $\rho_{tar}$ is the target density at the current position and $\rho_a$ is the estimated density for the particle (\eq \ref{eq:densest}). Higher $n$ means that fitting to the density is more important than balancing the particle distribution, that is, reducing $E_0$.  This is because a minimization in density error does not necessary mean a minimization in $E_0$. As such, density errors and particle distribution errors will drive the particles into a position that minimizes these errors, just as the case for SPH naturally\footnote{One could even set $n=0$, that is $P_a=1$ and relax the distribution, this would disregard the density estimate and instead assume an exact value of the density. This could be used if one evolves the density from the continuity equation $\frac{d\rho}{dt}$ for example, to minimize gradient errors. }.
$K_{global,i}$ is adjusted after each step following:
\begin{itemize}
    \item If the step leads to an overall worse mean $\sqrt{\sum \left(\frac{dv_i}{dt}\right)^2}$, then decrease $K_{global,i+1}=\frac{1}{d_1} K_{global,i}$.
    
    \item If the step leads to an overall better mean $\sqrt{\sum \left(\frac{dv_i}{dt}\right)^2}$, then increase $K_{global,i+1}=d_2 K_{global,i}$.
\\  
    \item Relaxation requires that $d_1 > d_2 > 1.0$. We set $d_1=1.8$ and $d_2=1.1$ during the global relaxation and $d_1=1.1$ and $d_2=1.002$ during local relaxation.
\\
    \item \textbf{The global relaxation stage:} Quickly reduces the step size to interparticle size to get a good gradient estimate of the current global density profile. The step size is then increased again to a factor of the initial ($C_{loop,n} K_{global,0}$), here we set $C_{loop,0}$ to 0.5 and reduce it by half each time $C_{loop,n}=0.5C_{loop,0}$ if there is no improvement to the global density error. If one starts from an IC that captures the global IC well, then this step is not necessary and only local relaxation is needed.
\\
    \item \textbf{The local relaxation stage:} When global relaxation stage is finished, the $K_{global,i}$ is near the average interparticle scale. From here on we slow down the rate of relaxation to ensure a more accurate local particle distribution. We stop the IC generator when $\max(\Delta r_a/h_a)< 2\cdot 10^{-3}$, at which point particle motions are negligible and we consider the distribution relaxed.    
\end{itemize} 
The initial value of $K_{global,i}$ is a parameter that can be chosen, but is by default set to half the system size, as this ensure that the desired density profile is obtained, regardless of the input IC. A much lower value can be used if one has an IC that is close to the desired density profile. 
\\ \\
Results from the IC generator in the case of the $\Delta\rho=8$, $n_x=64$, $N_{smooth}=64$ Kelvin-Helmholtz IC of \sect \ref{sec:reskelvin} can be seen in \fig\ref{fig:ICGEN}. From this we can see that we have a proper glass structure in the particle distribution that follows the analytical density profile very closely. As different relaxed glasses will be produced by different smoothing kernels and neighbour numbers, we run the relaxation process for each new kernel and neighbour number when generating ICs in this paper.
\subsection{Adjusting density bias}
\label{sec:densitybias}
\begin{figure}
    \centering
    \includegraphics[width=1.0\linewidth]{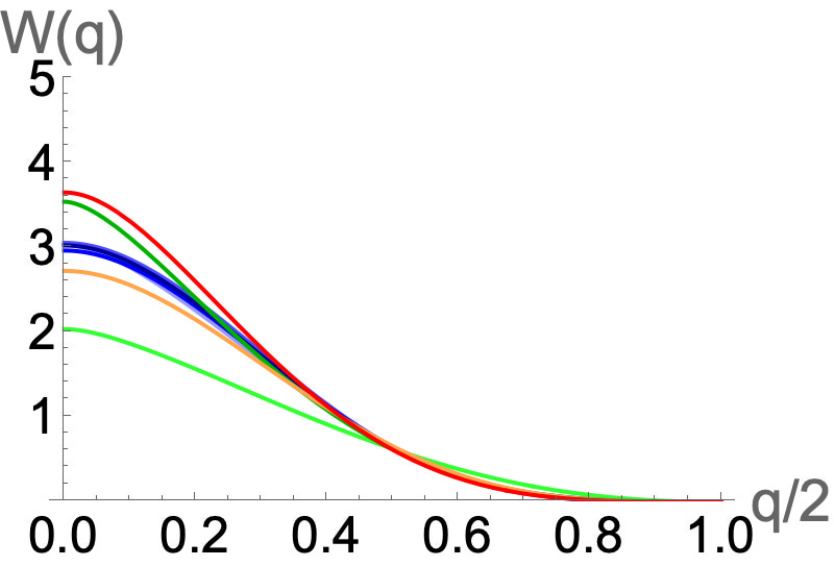}
    \includegraphics[width=1.0\linewidth]{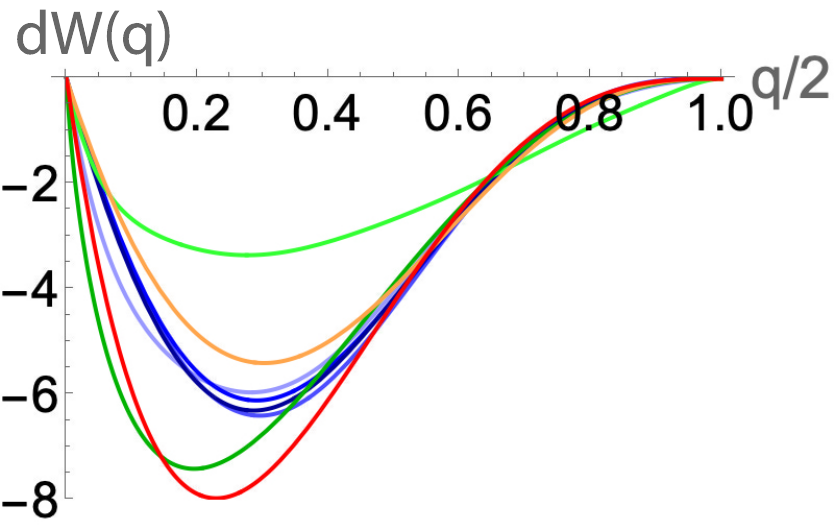}
    \includegraphics[width=1.0\linewidth]{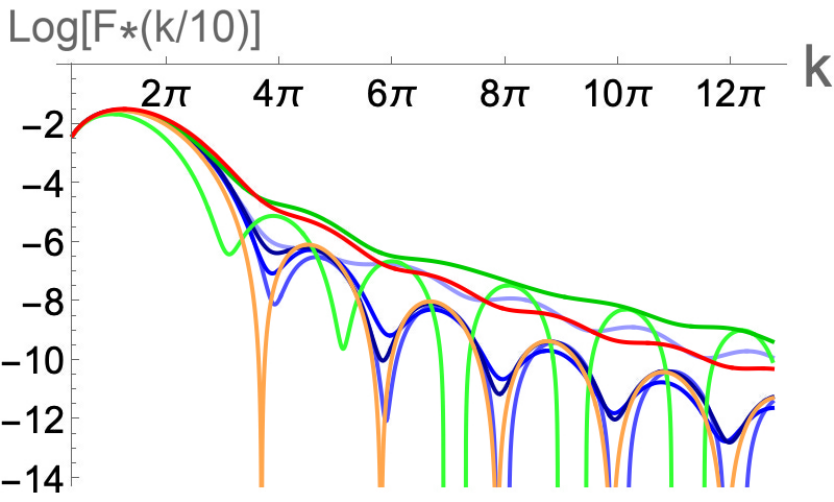}
    \includegraphics[width=1.0\linewidth]{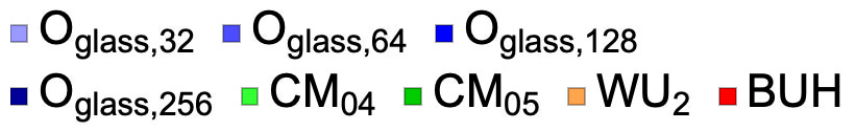}

    \caption{Top figure shows the optimized kernel functions for a glass distribution (\sect \ref{sec:optglass}) together with the missing Wendland kernels ($CM_{04},CM_{05}$), the Buhmann kernel ($BUH$) and the Wu kernel ($WU_2$). The middle figure shows the corresponding derivative of these kernel function and the bottom figure displays the Fourier transform of these kernels, scaled by $k/10$. The optimized kernels are quite similar, with some small variations being seen in the Fourier transform. The $O_{glass,32}$ kernel deviates the most from the other kernel, with less deep troughs in the Fourier transform. These kernel's generate excellent glass distributions but degrade significantly in dynamical simulations.}
    \label{fig:kernel1}
\end{figure}
Before we optimize our linear combined kernel, we adjust the self-bias of all our kernels. This mainly effects the density, giving us a corrected density estimate:
\begin{equation}
    \rho_{a}=\rho_{a,estimate}-\epsilon m_a W(0,h_a)
\end{equation}
We adjust $\epsilon$ so that the density of a relaxed hydrostatic glass is as close as possible to $\rho=1.0$ (given that $M_{tot}/V_{tot}=\rho=1.0$) for a wide range of neighbour numbers. A uniform glass produced by the IC generator will locally be independent of the resolution and only be a function of the number of neighbours and the given kernel. As such we can run low resolution cases with our IC generator and adjust the kernel bias until we produce a glass with a mean density error of around $\rho_{err}\approx 10^{-5}$, which is much less then the general SPH density noise. We run this for a wide range of neighbour numbers ($N_{smooth} \in \{16,24,32,48,64,96,128,256,512\}$). After 512 neighbours the kernel bias is set to 1. We can then interpolate between ranges using a linear piecewise function to correct the kernel:
\begin{equation}
    \epsilon(N_{smooth}) =\frac{ c_{i+1} - c_i}{N_{min,i+1}-N_{min,i}}(N_{smooth}-N_{min,i})+c_i
\end{equation}
Here $c_i$ is the $\epsilon$ value at $N_{smooth}=N_{min,i}$. This was performed for both GDSPH and ISPH. The coefficients and neighbour number can be found in Appendix \ref{sec:selfbias} in Table \ref{tab:biascoeff} for GDSPH and Table \ref{tab:biascoeff2} for ISPH.
\subsection{Optimization strategy}
\label{sec:optstrat}
In this method, we generate a new kernel by linear combination: 
\begin{equation}
\begin{split}
    W_{new,i} (\alpha_i,\alpha_{i+1}) &= \left(\alpha_i W_{new,i-1} +(1-\alpha_i)W_{i}\right)\alpha_{i+1} \\ 
    & +(1-\alpha_{i+1})W_{i+1} \\
\end{split}
\end{equation}
Here $\alpha_i$ and $\alpha_{i+1}$ are the coefficients that we optimize for. The process starts with $i=1$ and an initial kernel $W_{new,0}$. After determining $\alpha_i$, $\alpha_{i+1}$ by minimizing some cost function, we generate a new kernel ($W_{new,i}$). We then repeat this process with the newly generated kernel and two other kernels. This iterative process continues until we have used all the kernels that we wish to include in our linear combination. We use the differential evolution method to find the global minimum of the two parameters for each iteration. We set the initial parameter value to $(\alpha_i,\alpha_{i+1})=[0.0,0.0]$ and we set minimum and maximum parameters boundaries to be $bnds=[ (-3.0,3.0),(-3.0,3.0)]$.
\\ \\
After the optimization is complete we fit the linear-combination of kernels to a polynomial in the form:
\begin{align}
    P(x) = \sum_{i=0}^{8} c_i x^i\\
    W(q) = \sigma \left(1+q^2P(q)\right)
\end{align}
During the work of this paper we used two different cost functions and two different cases to optimize the kernel for, which is described in the sections below.
\subsubsection{Optimized for hydrostatic glass}
\label{sec:optglass}
\begin{figure}[h!]
    \centering
    \includegraphics[width=1.0\linewidth]{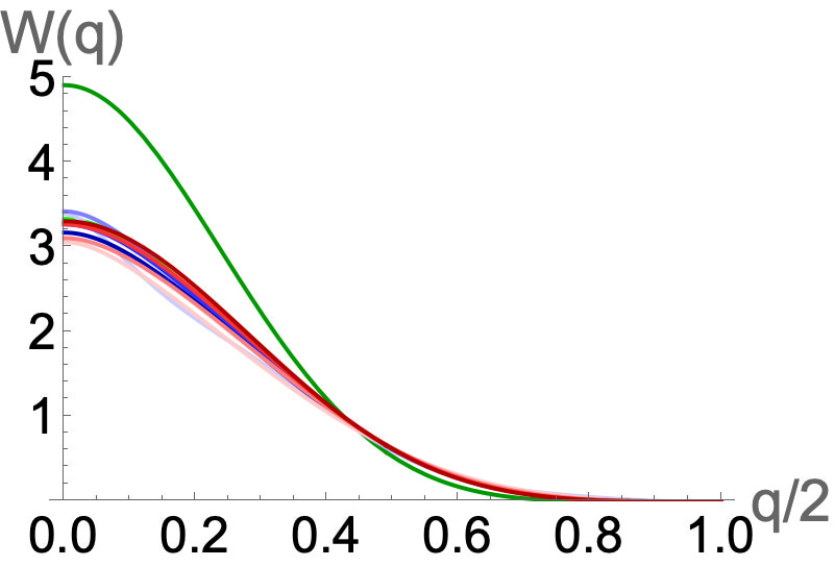}
    \includegraphics[width=1.0\linewidth]{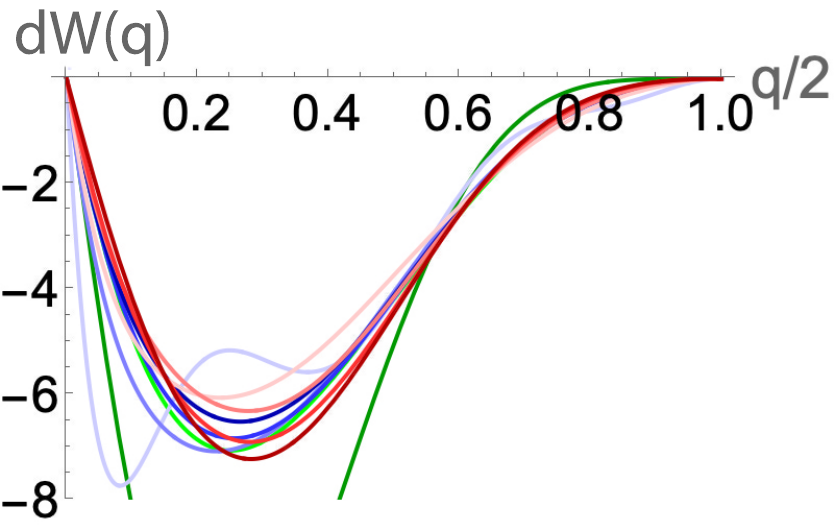}
    \includegraphics[width=1.0\linewidth]{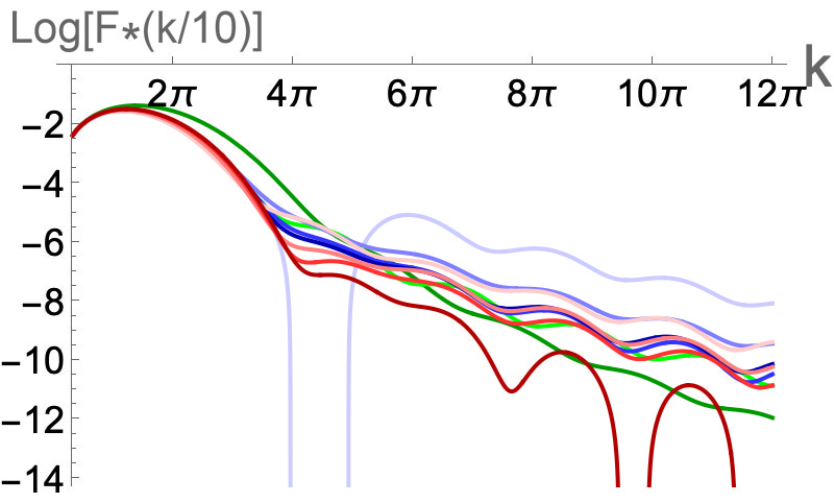}
    \includegraphics[width=1.0\linewidth]{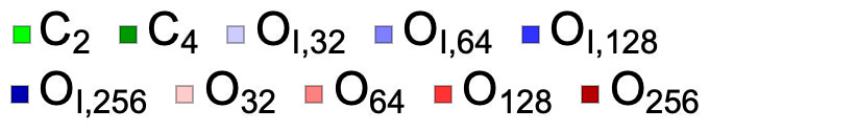}
    \caption{Top panel shows the optimized kernel functions for the Gresho-Chan vortex (\sect \ref{sec:optgresho}), for both GDSPH and ISPH together with the popular Wendland kernels ($C_2$,$C_4$). The middle figure shows the corresponding derivative of these kernel function and the bottom figure displays the Fourier transform of these kernels, scaled by $k/10$. The optimized kernel exhibit a clear trend in it's Fourier transform, where the kernels optimized for higher $N_{smooth}$ emphasizes lower power beyond $k>3\pi$. The general form of the optimized kernel is quite similar to the Wendland $C_2$ kernel. In general GDSPH produces lower power than ISPH in the high wave number regime for a given $N_{smooth}$. The $O_{I,32}$ kernel is the only optimized kernel that exhibits a negative Fourier transform at low wavenumbers. Among the other optimized kernels, only $O_{256}$ has a negative Fourier transform, but this occurs at higher wavenumbers and remains stable to pairing instability up to atleast $N_{smooth}<500$.
    }
    \label{fig:kernel2}
\end{figure}
Perhaps, one of the simplest cases to determine the quality of a smoothing kernel comes from it's ability to generate a hydrostatic glass that minimizes the leading errors of SPH. The hydrostatic glass models a homogeneous system in pressure equilibrium. A cubic box with length $L=1$ and total mass $M_{box}=1$ is setup. Particles are randomly distributed throughout the box. Due to the inherent re-meshing property of SPH, the particles will be reorganized to a glass-like structure by the low order errors (\eq \ref{eq:E0}, \eq \ref{eq:E1}). At some point the distribution will be organized into a relaxed glass, where the errors have been minimized (there is minimal/oscillatory change in the mean errors from this point onward). The distribution and the magnitude of the errors in this final relaxed state will depend on both the smoothing kernel, number of neighbours and the SPH method. The leading order error ($E_0$) will locally be independent of the resolution (for uniform density), so we use a relatively low resolution of $N=32^3$ particles. We hypothesized that the relaxed glass with the smallest errors/noise would provide the best result in dynamic cases as well. We defined our cost function for this case as:
\begin{equation}
    F=\delta\rho_{err}+E_0+0.25E_1
    \label{eq:costfunction}
\end{equation}
Here $E_0$ and $E_1$ are given by the norm of the error vectors defined in (\eq \ref{eq:E0}, \eq \ref{eq:E1}) respectively. $\delta\rho_{err}$ is given by:
\begin{equation}
\delta\rho_{err}=\frac{1}{N}\sum_i\lvert\rho_i-\rho_{mean}\lvert
\label{eq:denserr}.
\end{equation}
Each kernel has had its self-bias adjusted, so that $\rho_{mean}$ will be within $10^{-5}$ of $\rho=1.0$, for their respective glass and kernel (see \sect \ref{sec:densitybias}), such that we accurately measure the density noise. We only optimized this case with the GDSPH formula, and we refer to these kernels as $O_{glass,N_{smooth}}$ ($N_{smooth}=32,64,128,256$) for the rest of the paper. The resulting polynomial coefficients for the optimization are given in Table \ref{tab:glasspolycoeff}.
\\ \\
The kernel function, its derivative and its Fourier transform can be seen in \fig \ref{fig:kernel1} for the optimized hydrostatic glass kernels together with $WU_{2}$, $BUH$, $CM_{04}$, $CM_{05}$ kernels. All the optimized kernels for different $N_{smooth}$ look very similar to each other, being a bit steeper than the $WU_{2}$ kernels. However, some differences can be seen in the Fourier transform, where $O_{glass,32}$ has quite a flat trend (similar to $BUH$ kernel) , the higher $N_{smooth}$ kernels experience stronger oscillations (behavior somewhere in-between the $WU_2$ kernel and the $BUH$ kernel). Even though we allow for the linear combination to result in negative Fourier transforms, we only end up with one kernel that has negative Fourier transform in the high wave number regime ($O_{glass,64}$). The performance of these kernels are discussed more in the next section.

\subsubsection{Optimized for Gresho-Chan vortex}
\label{sec:optgresho}
\begin{figure*}[!h]
    \centering
    \includegraphics[width=1.0\linewidth]{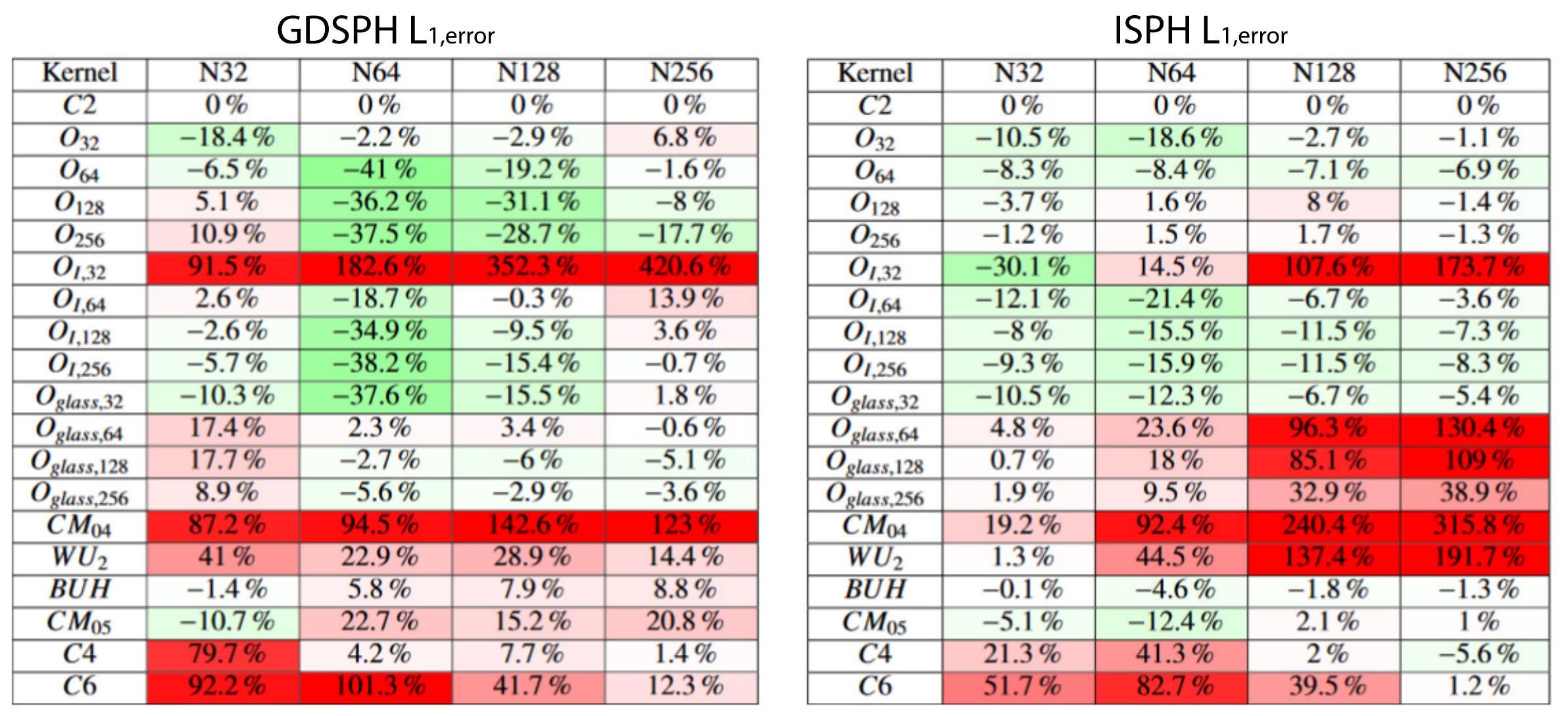}
    \caption{The tables above show the relative increase/decrease of the $L_{1,error}$ of the Gresho-chan vortex at $t=1$ and $n_x=64$ for different kernels in respect to the $L_{1,error}$ of the $C_2$ kernel. The N32, N64, N128, N256 refers to the number of neighbors $N_{smooth}=32,64,128,256$. The left table show simulations done with the GDSPH method and the right table show simulations done with the ISPH method. The $L_{1,error}$ of the C2 kernel for each $N_{smooth}$ is $L_{1,error}=1.49375\times10^{-1},9.57342\times10^{-2},4.70888\times10^{-2},2.99816\times10^{-2}$ for GDSPH and $L_{1,error}=9.34635\times10^{-2},4.66192\times10^{-2},2.64744\times10^{-2},2.22268\times10^{-2}$ for ISPH. We can see an improvement for the optimized kernel at a given $N_{smooth}$ across all $N_{smooth}$ for both methods, where the biggest improvement can be seen for the $N64$ and $N128$ for GDSPH and $N32$ and $N64$ for ISPH. Of the non-optimized kernels we can see that the $C_2$ kernel performs best, except for $N_{smooth}=32$ (GDSPH), $N_{smooth}=32,64$ (ISPH), where $CM_{05}$ performs better and $C_4$ performs slightly better at $N_{smooth}=256$ (ISPH).}
    \label{fig:greshoL1t1}
\end{figure*}

Initially, we assumed that a smoothing kernel with minimal errors in its final relaxed glass state would also yield the best results in dynamic simulations, but this turned out to not be the case. As different kernels have different sensitivity when it's glass gets disturbed. This was clearly seen when testing out the Gresho-Chan vortex test case. Here, even though the density is uniform, the particle distribution is continuously disturbed by a shear flow.
\\ \\
The Gresho-Chan vortex simulates an inviscid fluid vortex in force equilibrium \citep{1990IJNMF..11..621G}. The setup is as follows: We setup a 3D thin periodic box with length $L=(1,1,2\frac{\sqrt{6}}{n_x})$, here $n_x$ is the average number of particles in the $x$ direction, and the z boundary is set to be roughly 24 particle spacings. The pressure profile is setup following:
\begin{equation}
P(r) = 
\begin{cases}
5+ \frac{25}{2}r^2 & \text{for } 0 \leq r < 0.2 \\
9 + \frac{25}{2}r^2 - 20r + 4\ln 5r & \text{for } 0.2 \leq r < 0.4 \\
3 + 4\ln r & \text{for } r \geq 0.4
\end{cases}
\end{equation}
And the velocity profile follows:
\begin{equation}
v_\theta(r) = 
\begin{cases}
5r & \text{for } 0 \leq r < 0.2 \\
2 - 5r & \text{for } 0.2 \leq r < 0.4 \\
0 & \text{for } r \geq 0.4
\end{cases}
\end{equation}
Even though conservation of angular momentum within SPH is near exact, the test is challenging for SPH due to the artificial viscosity induced by the particle noise and the errors in the linear gradient (\eq \ref{eq:E1}). The artificial viscosity will cause angular momentum transport, which will lead to mass accretion towards the centre and a loss in kinetic energy. Increasing the number of neighbours \citep{2012MNRAS.422.3037R,2015MNRAS.448.3628R}, using artificial viscosity switches \citep{2010MNRAS.408..669C} and doing linear gradient corrections \citep{2012A&A...538A...9G,2016ApJ...831..103V,2017A&A...606A..78C} all improve the convergence of this case.
\\ \\
While testing the Gresho-Chan vortex we found that only $O_{glass,32}$ outperformed the other kernels for the $N_{smooth}$ it had been optimized for, while all the other $O_{glass}$ kernels performed rather badly (can be seen in \fig \ref{fig:greshoL1t1}). Similarly, we saw bad result for all kernels that exhibit strong oscillatory behavior in their Fourier transform, like the $WU_2$ kernel (\fig \ref{fig:kernel1}). Therefore we decided to additionally optimize kernels based on results from the Gresho-Chan vortex following the procedure laid out in the previous sections. The main difference is the cost function, which we set to be the $L_{1,error}$ at $t=1$:
\begin{equation}
    F=L_{1,error}(t=1)
\end{equation}
\begin{equation}
L_{1,error} = \frac{1}{N} \sum_{i=1}^{N} \lvert v_{\phi,i}^{a}(r)-v_{\phi,i}^{d}(r)\lvert
\label{eq:L1}
\end{equation}
Here $v_{\phi,i}^{a}$ stand for the analytical and $v_{\phi,i}^{d}$ the result from the simulation. Similarly to all other setups, we use the IC generator (\sect \ref{sec:ICgen}) to generate an initial glass for each kernel and neighbour number before each run. We optimize both GDSPH and ISPH for this case; here the kernels optimized with GDSPH are given by $O_{N_{smooth}}$ and with ISPH $O_{I,N_{smooth}}$, which was done for $N_{smooth}=32,64,128,256$. The kernels and the resulting coefficients that were used for the optimization are given in Table \ref{tab:greshopolycoeff} and Table \ref{tab:greshokernel}.
\\ \\
The kernel function, its derivative and its Fourier transform can be seen in \fig \ref{fig:kernel2} for the optimized kernels together with $C_2$ and $C_4$ kernels. All the resulting kernels have similar amplitude and form as the Wendland $C2$ kernel (except for $O_{I,32}$), but with a minima in its derivative being a bit offset from the $C2$ kernel. Looking at the Fourier transform we can see quite a clear trend with both the $O_{N_{smooth}}$ and $O_{I,N_{smooth}}$ as we increase $N_{smooth}$. Both $O_{32}$ and $O_{I,32}$ drop the fastest within $k \propto 0-3 \pi$ but remain higher than the other kernels as we go to larger wave numbers. As we increase $N_{smooth}$ we have a higher power between $k \propto 0-3 \pi$ but lower for higher wave numbers. We have two resulting kernels with regions of negative Fourier transform. The $O_{I,32}$ kernel, which goes negative in the region $k \propto 4-5 \pi$, and becomes susceptible to the pairing instability as we go to higher $N_{smooth}$ (we see instability at $N_{smooth}=128$ and $N_{smooth}=256$). The $O_{256}$ also has regions of negative Fourier transform at much higher wave numbers ($k \propto 10-12 \pi$), however, we do not see any sort of instability occurring in this kernel, at least up to $N_{smooth}=512$. The performance of these kernels are discussed more in \sect \ref{sec:results}.
\section{Results}
\label{sec:results}
\begin{figure}[]
    \centering
    \includegraphics[width=1.0\linewidth]{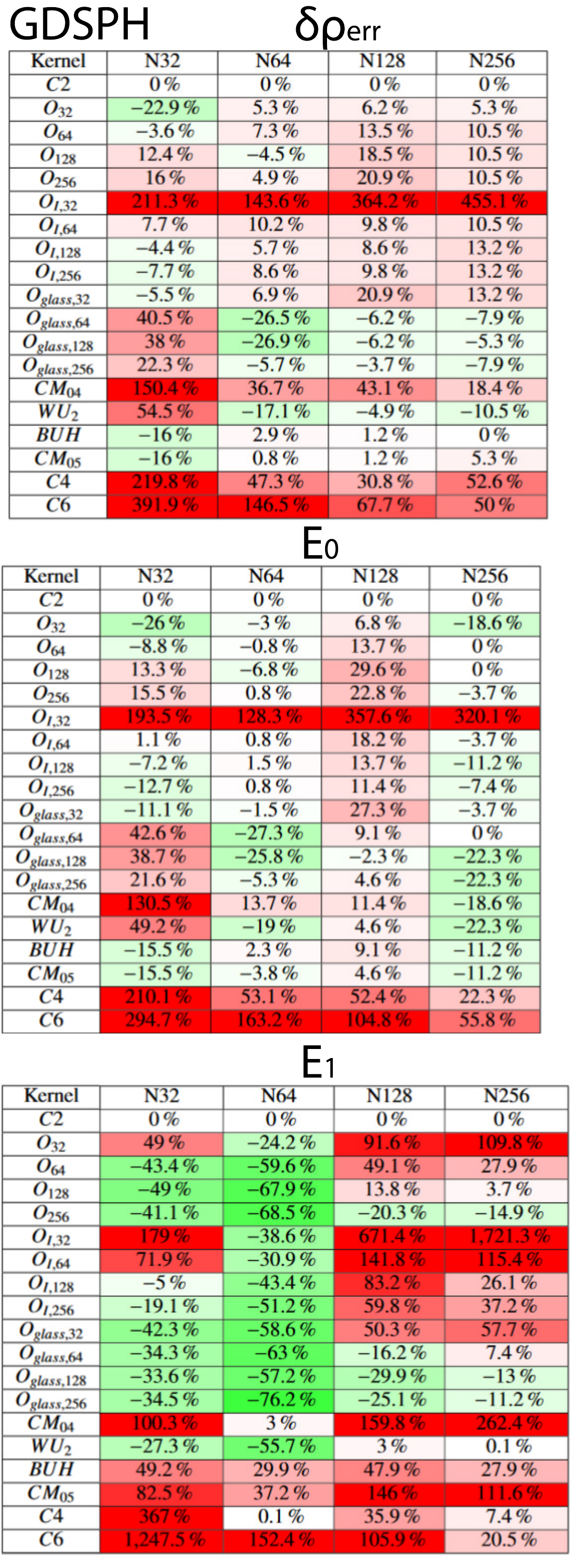}
    \caption{The tables above show the relative increase/decrease of the $\delta \rho_{err}$ (top), $E_0$ (middle) and $E_1$ (bottom) of the hydrostatic glass when relaxed for different kernels in respect to the corresponding error of the $C_2$ kernel. The simulations here was performed with only the GDSPH method. The N32, N64, N128, N256 refers to the number of neighbors $N_{smooth}=32, 64, 128, 256$ respectively. The $\delta \rho_{err}$, $E_0$ and $E_1$  of the C2 kernel for each $N_{smooth}$ is $\delta \rho_{err}=5.539\times 10^{-3},3.747\times 10^{-3},1.243\times 10^{-3},5.860\times 10^{-4}$, $E_0=2.763\times 10^{-3},2.024\times 10^{-3},6.797\times 10^{-4},4.253\times 10^{-4}$ and $E_1=2.210\times 10^{-2},1.484\times 10^{-2},2.557\times 10^{-3},8.305\times 10^{-4}$. The optimized kernels can in general be seen to drastically reduce the $E_1$ error relative to the C2 kernel. For the glass optimized kernel's we can also see a general decrease in both $\delta \rho$ and $E_0$ for their given $N_{smooth}$
    }
    \label{fig:glassresults}
\end{figure}
\begin{figure*}[]
    \centering
    \includegraphics[width=1.0\linewidth]{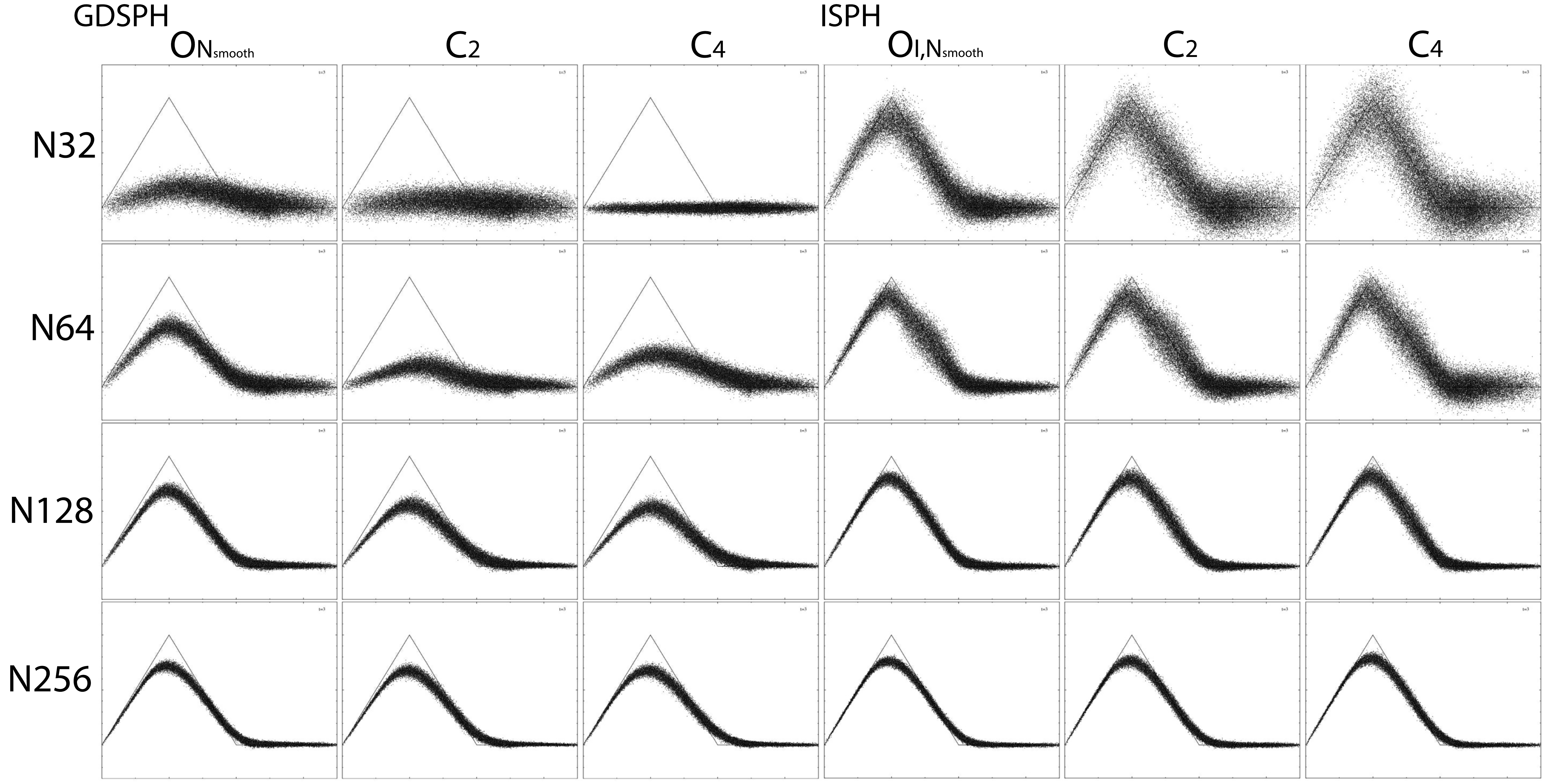}
    \caption{The azimuthal velocity in the radial direction of the Gresho-Chan vortex at $t=3$. Here the resolution is $n_x=64$ and simulation was performed with GDSPH (left 3 columns) and ISPH (right 3 columns). We compare the optimized kernel at a given $N_{smooth}$ to the popular Wendland $C_2$ and $C_4$ kernels, and vary the number of neighbours in each row ($N_{smooth}=32,64,128,256$), referred here to N32, N64, N128, N256. The solid black line show the analytical solution and the black dots show the results from the simulation. We can see a significant improvement in the optimized kernels for the GDSPH method for $N_{smooth}=32,64,128$, where the vortex structure becomes more prounounced. Overall less noise can be seen for all $N_{smooth}$. The ISPH method captures the overall vortex structure well for all $N_{smooth}$, due to exact linear gradients. But we can see significant improvement in the noise level of the optimized kernels, compared to the $C_2$ and $C_4$ kernels.}
    \label{fig:greshoazi}
\end{figure*}
\begin{figure*}[]
    \centering
    \includegraphics[width=0.41\linewidth]{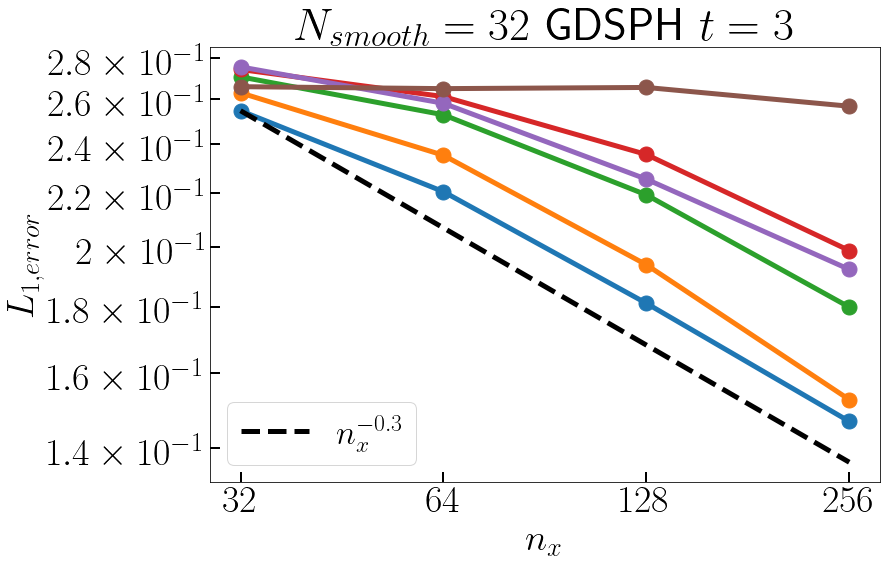}
    \includegraphics[width=0.41\linewidth]{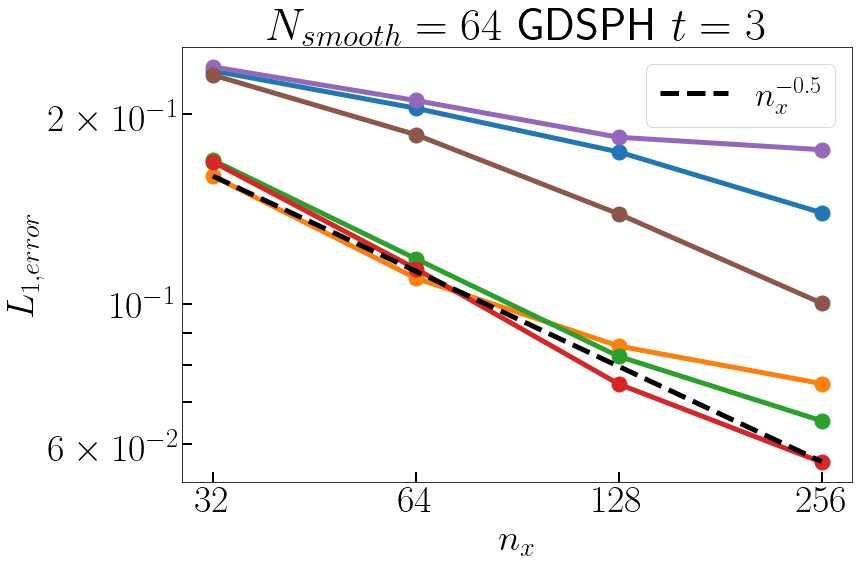}
    \includegraphics[width=0.41\linewidth]{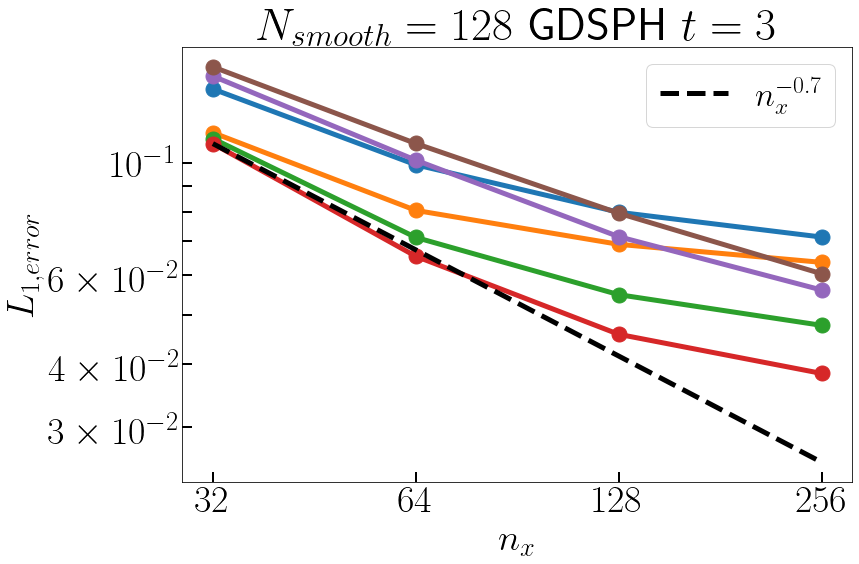}
    \includegraphics[width=0.41\linewidth]{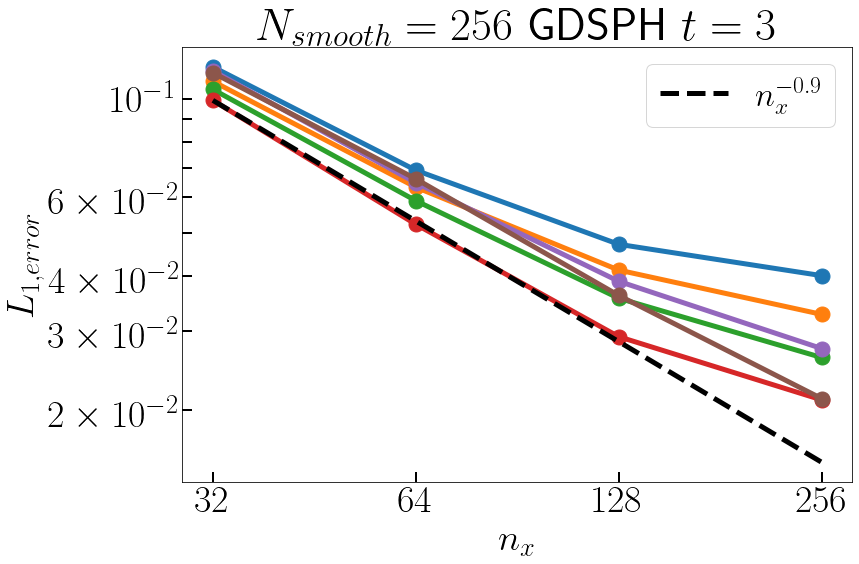}
    \includegraphics[width=0.7\linewidth]{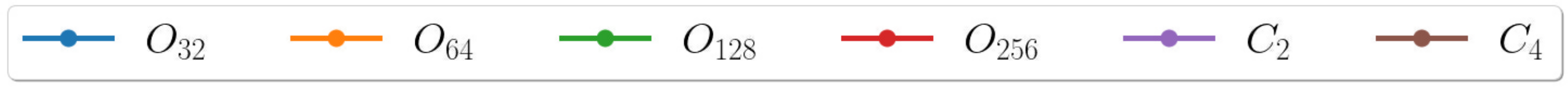}
    \caption{Convergence study of the Gresho-Chan vortex with the GDSPH method for $N_{smooth}=32,64,128,256$. Here we plot the $L_{1,error}$ over resolution elements in x direction ($n_x$) at $t=3$. Significant improvements can be seen for all optimized kernels compared to the Wendland C2 and C4 kernels. The kernels were optimized in respect to an earlier time $t=1$ and we can see that $O_{256}$ have an improved result over $O_{128}$ for $N_{smooth}=128$ at this later time, even beating out $O_{64}$ for $N_{smooth}=64$ at higher resolution ($n_x>128$). The $C4$ kernel exhibits slightly better convergence slope at high $N_{smooth}$ and $n_x>128$, likely due to it's higher order properties (less $O(h^2)$ errors).}
    \label{fig:GDSPHCONV}
\end{figure*}
\begin{figure*}[]
    \centering
    \includegraphics[width=0.41\linewidth]{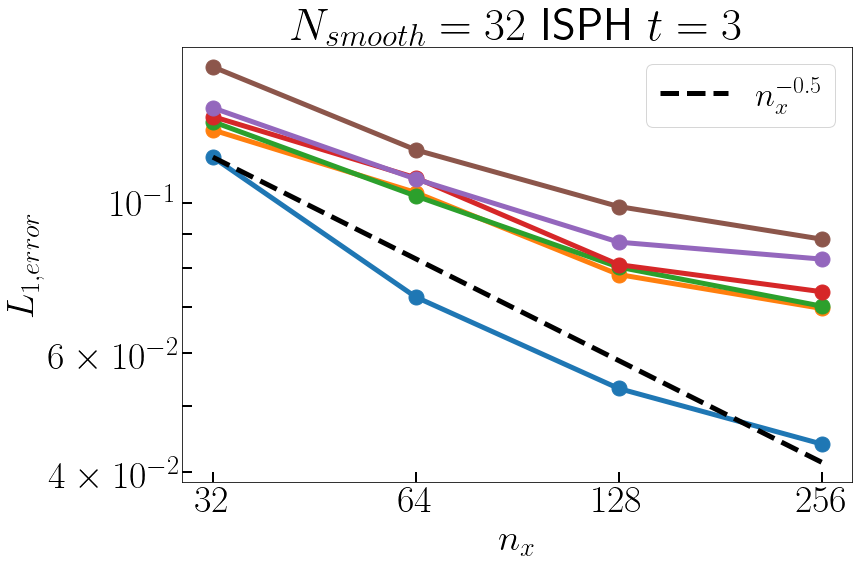}
    \includegraphics[width=0.41\linewidth]{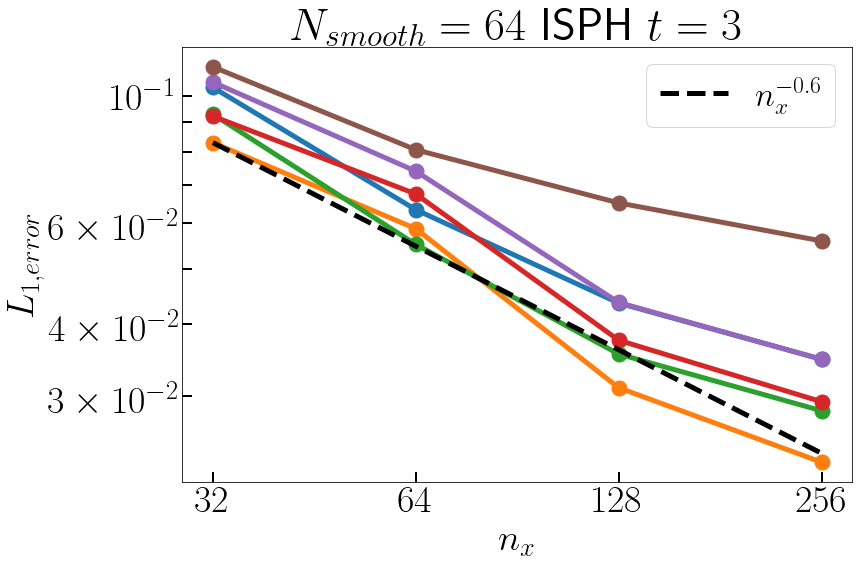}
    \includegraphics[width=0.41\linewidth]{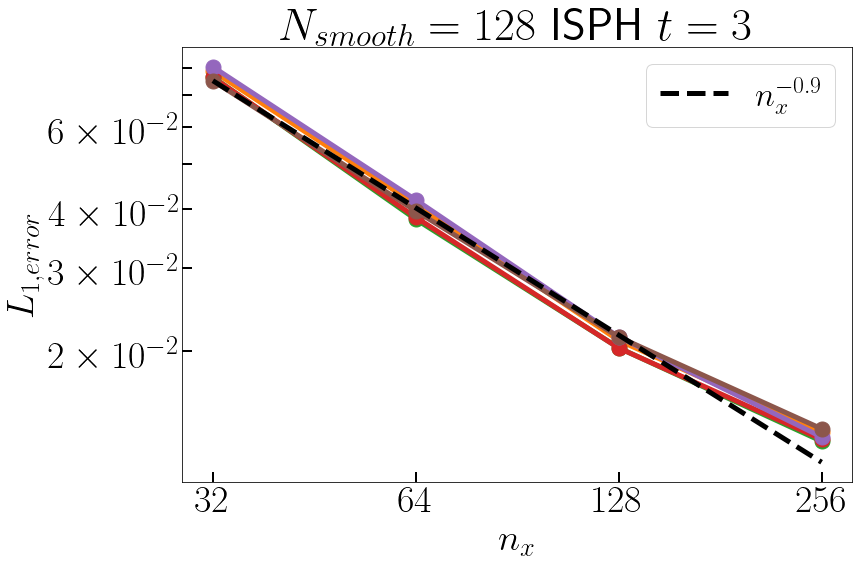}
    \includegraphics[width=0.41\linewidth]{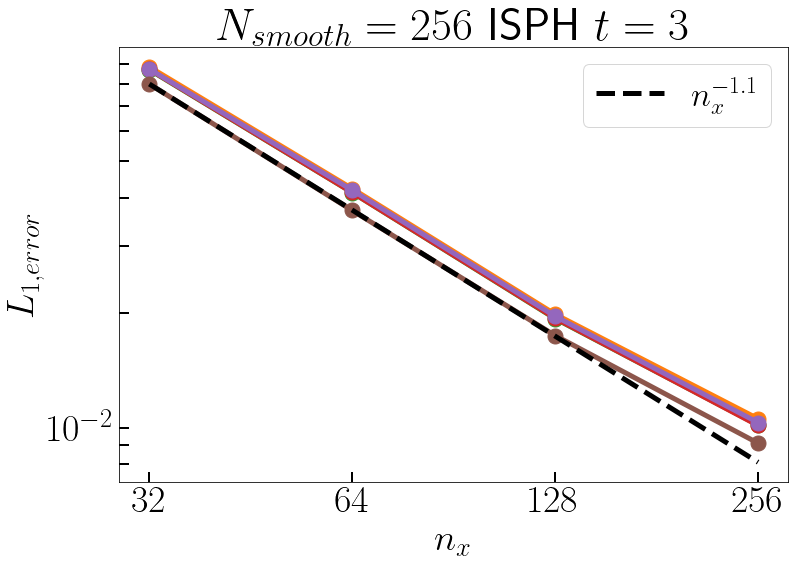}
     \includegraphics[width=0.7\linewidth]{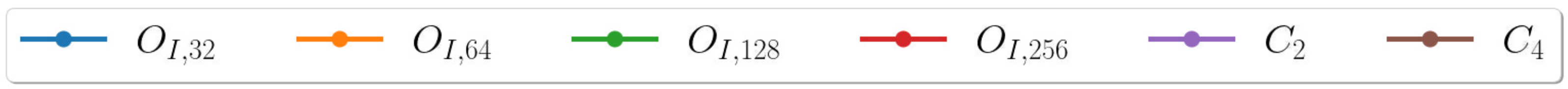}
    \caption{Convergence study of the Gresho-Chan vortex with the ISPH method for $N_{smooth}=32,64,128,256$. Here we plot the $L_{1,error}$ over the resolution elements in the $x$ direction ($n_x$) at $t=3$. Significant improvements can be seen for the optimized kernels at $N_{smooth}=32,64$ compared to the Wendland C2 and C4 kernels. While at higher $N_{smooth}$, there is much less difference between all the kernels. The kernels were optimized in respect to an earlier time $t=1$, and we can see that $C_4$ performs better at this later time for $N_{smooth}=256$ than all the optimized kernels at all resolutions. This is surprising as $E_0$ errors of the optimized kernels are much lower than the $C_4$ kernel (see \fig~\ref{fig:E0N256}), which might indicate that long term behavior at this $N_{smooth}$ is heavily determined by the higher order errors ($O(h^2)$).}
    \label{fig:ISPHCONV}
\end{figure*}
\begin{figure}[]
    \centering
    \includegraphics[width=\linewidth]{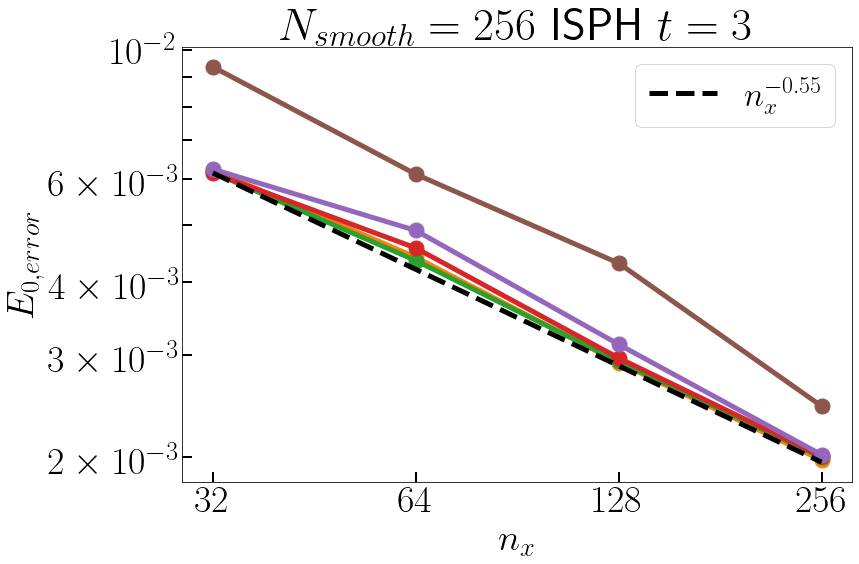}
    \includegraphics[width=0.6\linewidth]{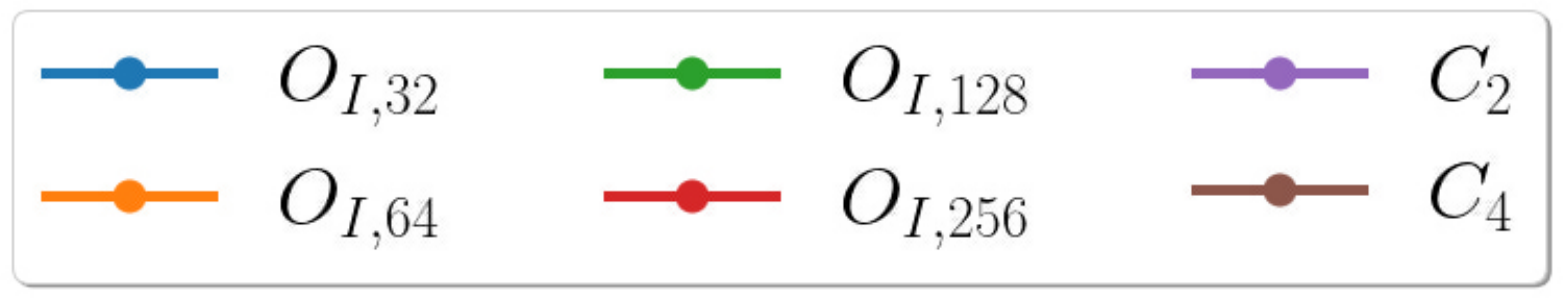}
    \caption{Convergence study of the Gresho-Chan vortex with the ISPH method for $N_{smooth}=256$. Here we plot the $E_0$ over resolution elements in the $x$ direction ($n_x$) at $t=3$. We can see much lower zeroth order errors for the optimized kernel and $C_2$ compared to the higher order $C_4$ kernel. Note here that even though being zeroth order (dependent on local particle order) the zeroth order decreases as we increase resolution ($n_x^{-0.55}$), due to the local particle distribution being more uniform.}
    \label{fig:E0N256}
\end{figure}
In the following section we discuss the performance of these optimized smoothing kernels on a select number of test cases. First we go through the hydrostatic glass and the Gresho-Chan vortex, both of which we outlined and used for optimization of the kernels in \sect \ref{sec:optglass} and \sect \ref{sec:optgresho}. We then go through two additional tests to see how well the optimized kernels perform in general by testing the Kelvin-Helmholtz instability for two different density contrasts ($\Delta \rho = 2, \Delta \rho = 8$) and the classic Sod shocktube test. The hydrostatic glass and Gresho-Chan vortex are run for all the optimized kernels. The ones optimized to hydrostatic glass with GDSPH: $O_{glass,32},O_{glass,64},O_{glass,128},O_{glass,256}$. The ones optimized for Gresho-Chan vortex with GDSPH $O_{32},O_{64},O_{128},O_{256}$ and with ISPH $O_{I,32},O_{I,64},O_{I,128},O_{I,256}$. We also compare with a select number of kernel's mentioned in \sect \ref{sec:kernelprop} ($CM_{04}, WU_2, BUH, CM_{05}, C2, C4, C6$). This is done for $N_{smooth}=32, 64, 128, 256$. For the Kelvin-Helmholtz instability and the Sod shocktube we compare the best performing kernel at a given $N_{smooth}$ with the popular Wendland C2 and C4 kernel, for both GDSPH and ISPH.
\subsection{Hydrostatic glass}
\label{sec:reshydroglass}
The setup of this test case is outlined in \sect \ref{sec:optglass}. The system is evolved for a long enough time to reach a relaxed glass state. The quality of the resulting glass is quantitied with the SPH errors ($E_0$ \eq \ref{eq:E0} $E_1$ \eq \ref{eq:E1}) and the density error (\eq \ref{eq:denserr}). These are given for all kernels and chosen neighbour numbers in \Fig~\ref{fig:glassresults}. While the kernel's optimized for the hydrostatic glass perform in average the best considering the cost function (\eq \ref{eq:costfunction}), we can see that it loses out to some other kernels for specific errors. All the GDSPH optimized kernels for the Gresho-Chan vortex perform fairly well, similar to the $BUH$, $CM_{05}$ and $C_2$ kernel for the $\delta \rho_{err}$ and $E_0$ error. The $O$ kernels does in general have smaller $E_1$ errors than the standard kernels. The only exception is the $O_{32}$ kernel that has higher $E_1$ error but has the smallest density and $E_0$ error out of all the kernels for $N_{32}$. The $WU_2$ kernel performs very well for higher neighbour numbers and is close to the results of the optimized kernel $O_{glass,128}$ and $O_{glass,256}$. The biggest improvement that can be seen when comparing the $O_{glass}$ and $O$ kernels with the Wendland $C2$ kernels, is in the $E_1$ errors that are significantly reduced. We can see that the $O_{I,32}$ kernel performs very badly when relaxing to a hydrostatic glass with the GDSPH gradient operator, even at $N_{smooth}=32$. This is, however, not the case when relaxing with ISPH at $N_{smooth}=32$, indicating that glass relaxation is different when using ISPH.
\subsection{Gresho-Chan vortex}
\label{sec:resgresho}
The setup of this test case is outlined in \sect \ref{sec:optglass}. We run the Gresho-Chan vortex up to $t=3$. We also perform a convergence study, to see if the optimized kernels still perform well as we increase/decrease the resolution, as the optimization was performed at $n_x=64$ resolution. For this convergence study we only compare with the Wendland $C_2$ and $C_4$ kernels. The optimization of the $O_{Nsmooth}$ kernels was done at $t=1$, we can see the result of the $L_{1,error}$ (see \eq \ref{eq:L1}) at $t=1$ in \fig \ref{fig:greshoL1t1} for both GDSPH and ISPH. From this table we can see that the optimized kernels for the Gresho-Chan vortex perform the best for their given $N_{smooth}$. The difference in $L_{1,error}$ is more pronounced for lower $N_{smooth}$ and for the GDSPH method compared to the ISPH method. A comparison of the azimuthal velocity structure ($n_x=64, t=3$) between the optimized kernel and the Wendland $C_2$ and $C_4$ kernel can be be seen in \Fig~\ref{fig:greshoazi} for GDSPH and ISPH. For GDSPH we can see that the optimized kernel mitigates the decay of the vortex structure (in particular for $N_{smooth}=64$). For ISPH the effect of the optimized kernels can mainly be seen as less noise in the vortex structure, while the general decay is similar across all the kernels (as this is highly dependent on the linear gradient estimate). A convergence study at $t=3$ of the $L_{1,error}$ can be seen in \Fig~\ref{fig:GDSPHCONV} for GDSPH and \Fig~\ref{fig:ISPHCONV} for ISPH. Looking at the convergence study, we can see that the improvement of the optimized kernels in general stays true for higher and lower resolution.
\\ \\
Looking at the GDSPH results, we can see that while the $O_{128}$ kernel proved to be the best at $t=1$, this is no longer true at $t=3$ and $N_{smooth}=128$, where the $O_{256}$ performs better across all resolutions at this time. And looking at the result at the earlier time of $t=1$, the $O_{128}$ just barely performed better at $N_{smooth}=128$ $n_x=64$, but worse than $O_{256}$ for higher resolution. We can also see that the $O_{256}$ performs better than the $O_{64}$ at $N_{smooth}=64$ for the higher resolutions($n_x=128,256$) at both $t=1$ and $t=3$. The improvement seen in $N_{smooth}=64$ for GDSPH is the largest improvement that we see in across all our Gresho-Chan cases, where the optimized kernels effectively half the $L_{1,error}$ compared to the Wendland $C_2$ and $C_4$ kernels, resulting in errors similar to $N_{smooth}=128$ errors of the Wendland kernels (at both $t=1$ and $t=3$). This is quite clearly seen in \Fig~\ref{fig:greshoazi}, when comparing the $O_{64}$ $N_{smooth}=64$ vortex structure with the $N_{smooth}=128$, $C_2$ and $C_4$ vortex structure. A quite surprising result is that the $C_4$ kernel performs better than the $C_2$ kernel for $N_{smooth}=64$, while performing worse for $N_{smooth}=32,128$, where at $N_{smooth}=32$ it performs drastically worse. For $N_{smooth}=32$, we can see that the $C_2$ kernel performs relatively well, but with $O_{32}$ being about a $20-30\%$ improvement at $t=1$ and the only kernel with some sort of vortex structure at $t=3$ (as can be seen in \Fig~\ref{fig:greshoazi}).
\\ \\
Looking at the result from ISPH, we can see a much less difference between all the different kernels, especially at high neighbour number ($N_{smooth}=128,256$). At $t=3$ and $N_{smooth}=256$ we can see that the $C_4$ kernel actually performs better than all other kernels, including the optimized kernel $O_{I,256}$. This is quite surprising, as the $O_{I,256}$ performs better at earlier times, and looking at the $E_0$ errors at $t=3$ in \Fig~\ref{fig:E0N256}, we can see that they are higher for $C_4$ compared to all other kernels. This likely means that $C_4$ has smaller higher order errors, compared to the other kernels. The $O_{I,32}$ kernel performs significantly better than all the other kernels at $N_{smooth}=32$ and is the only optimized kernel with negative Fourier transform at relatively low wave number, which means that it becomes unstable to pairing instability at higher neighbor numbers (above $N_{smooth}=64$). The Wendland $C_4$ kernel is by far the worst kernel at lower neighbour numbers ($N_{smooth}=32,64$) for ISPH. For $N_{smooth}=64$ at both $t=1$ and $t=3$, the $O_{I,64}$ produce a significant improvement compared to the Wendland kernels. 
\\ \\
The convergence scaling seems to be similar between kernels, but with some experiencing a flattening at higher resolution. The $C_4$ kernel seems to scale a bit better than other kernels at higher resolution, though still being worse in general. The scaling will also depend on the given dissipation scheme/parameters used. We have also run the Gresho-Chan vortex with different dissipation parameters ($\alpha=0.01,0.05,0.1$) and the relative results seen in this section remains the same.
\subsection{Kelvin Helmholtz}
\label{sec:reskelvin}
\begin{figure*}[]
    \centering
    \includegraphics[width=0.49\linewidth]{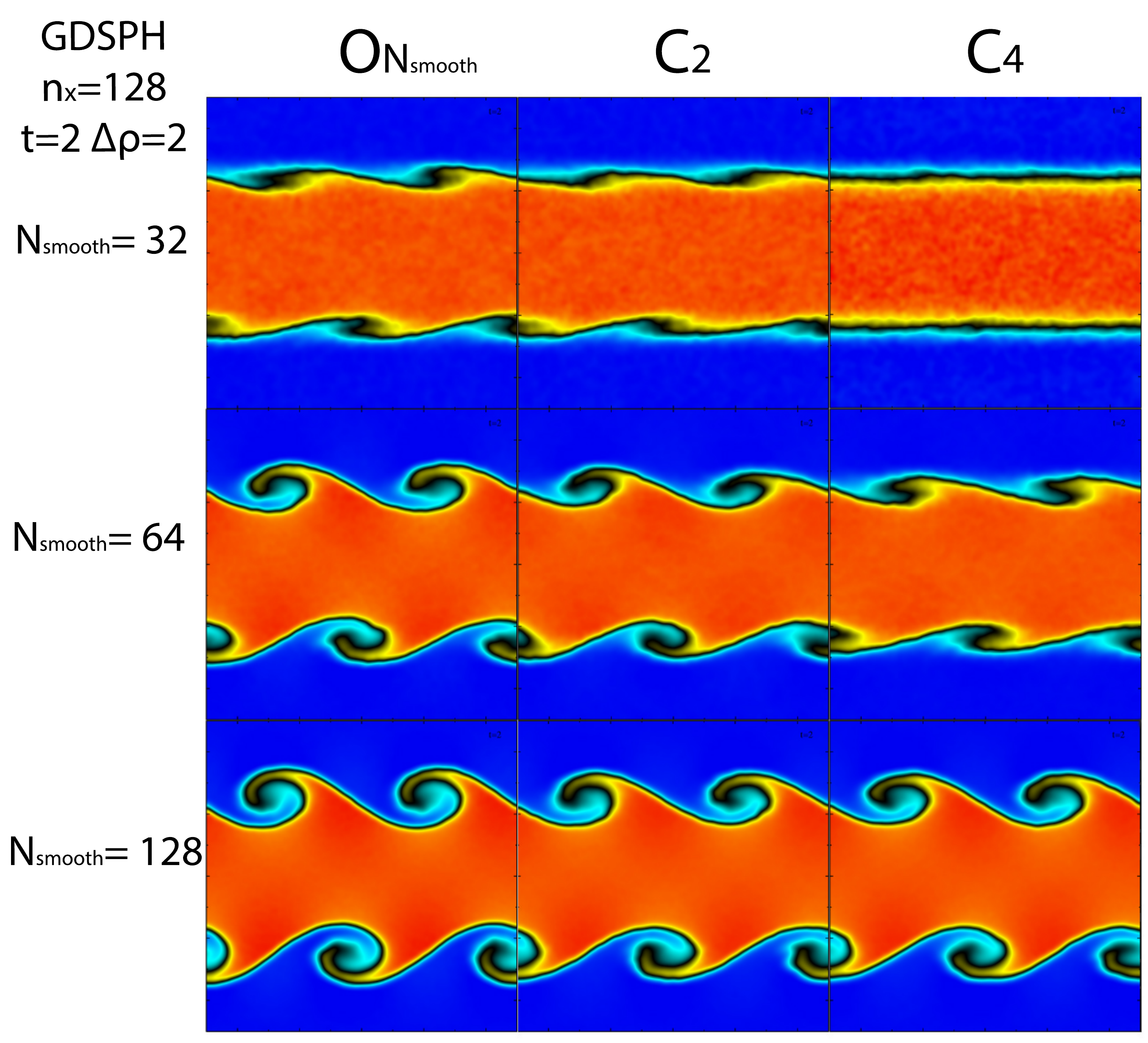}
    \includegraphics[width=0.49\linewidth]{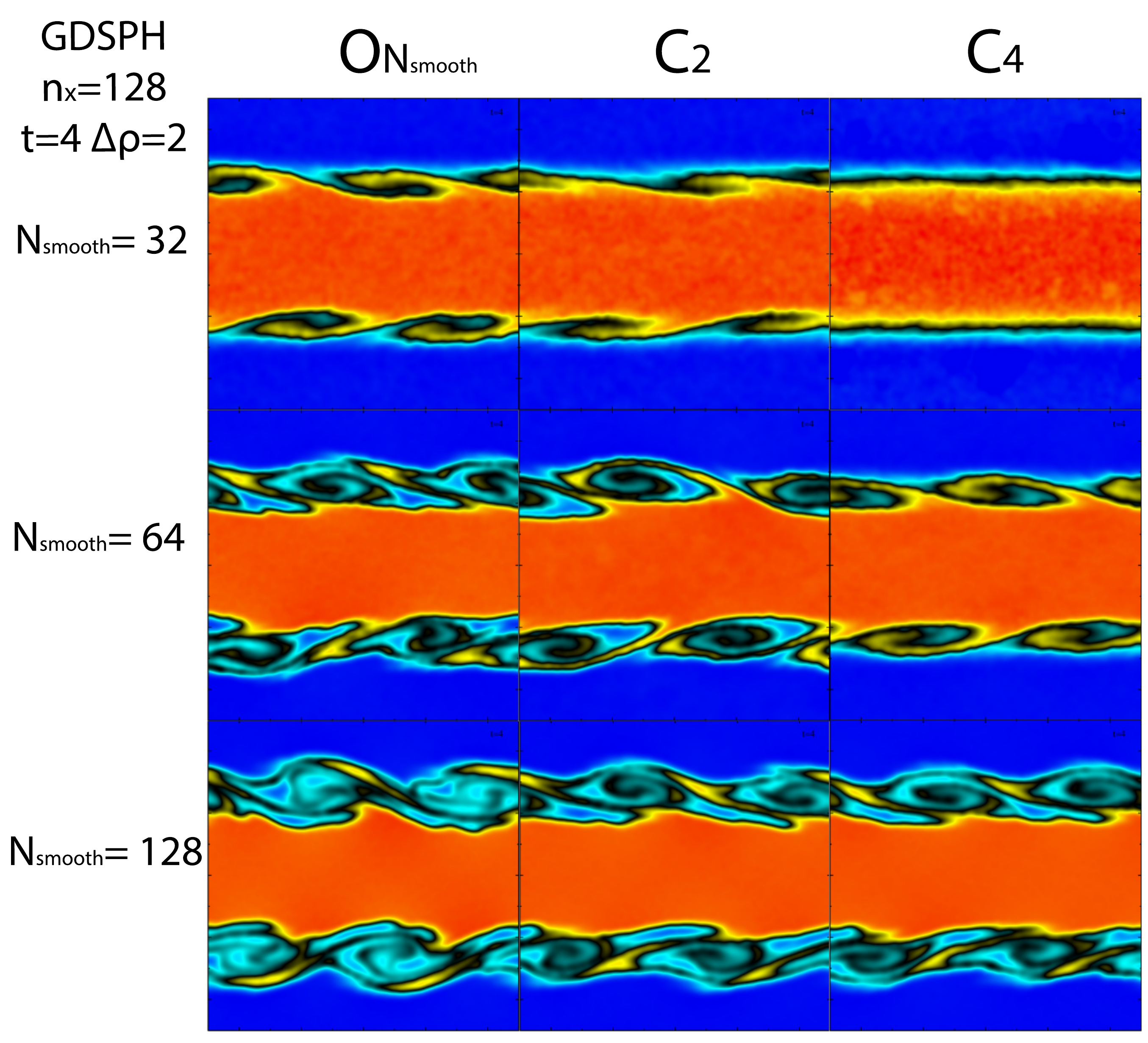}
    \includegraphics[width=0.49\linewidth]{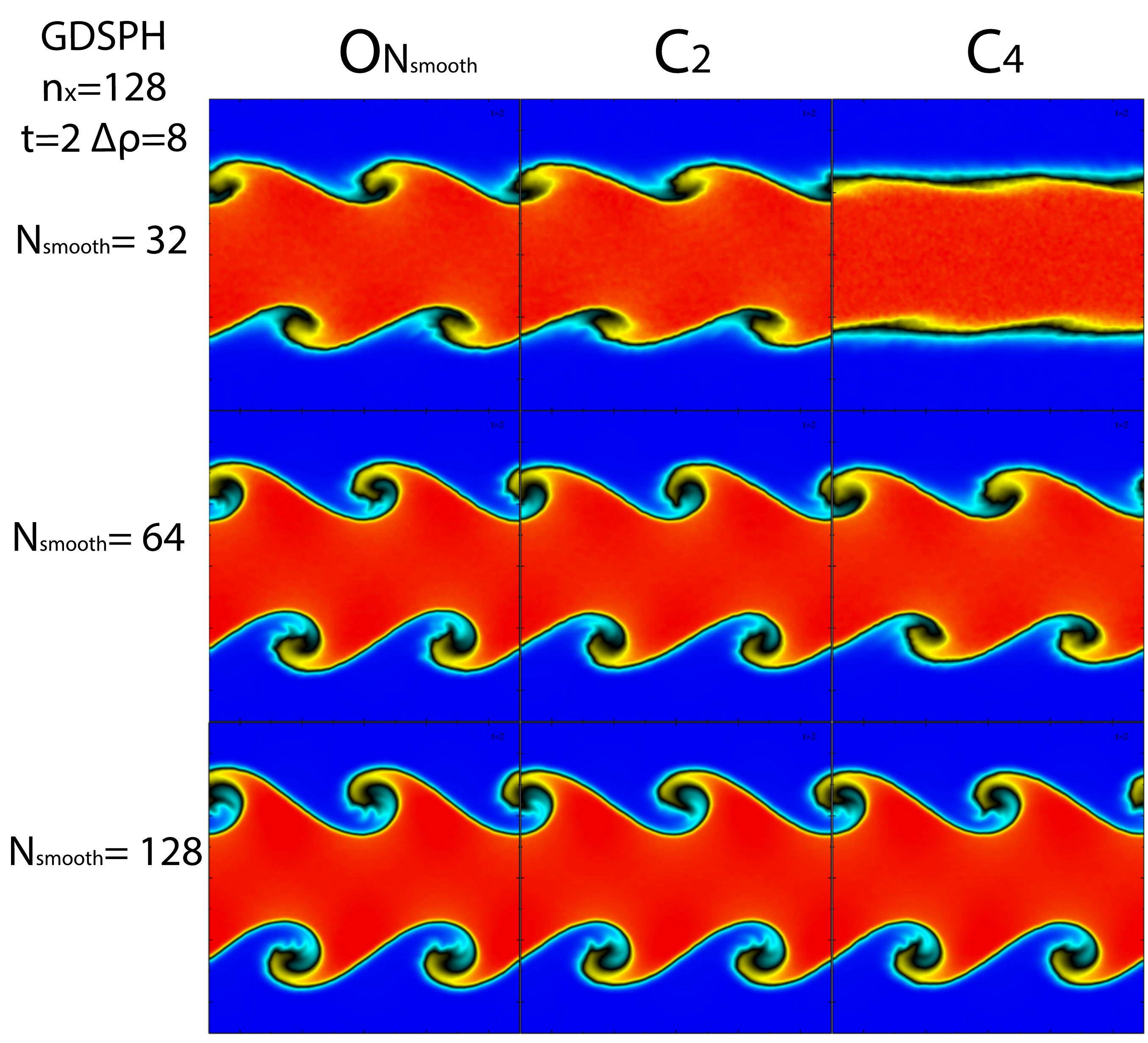}
    \includegraphics[width=0.49\linewidth]{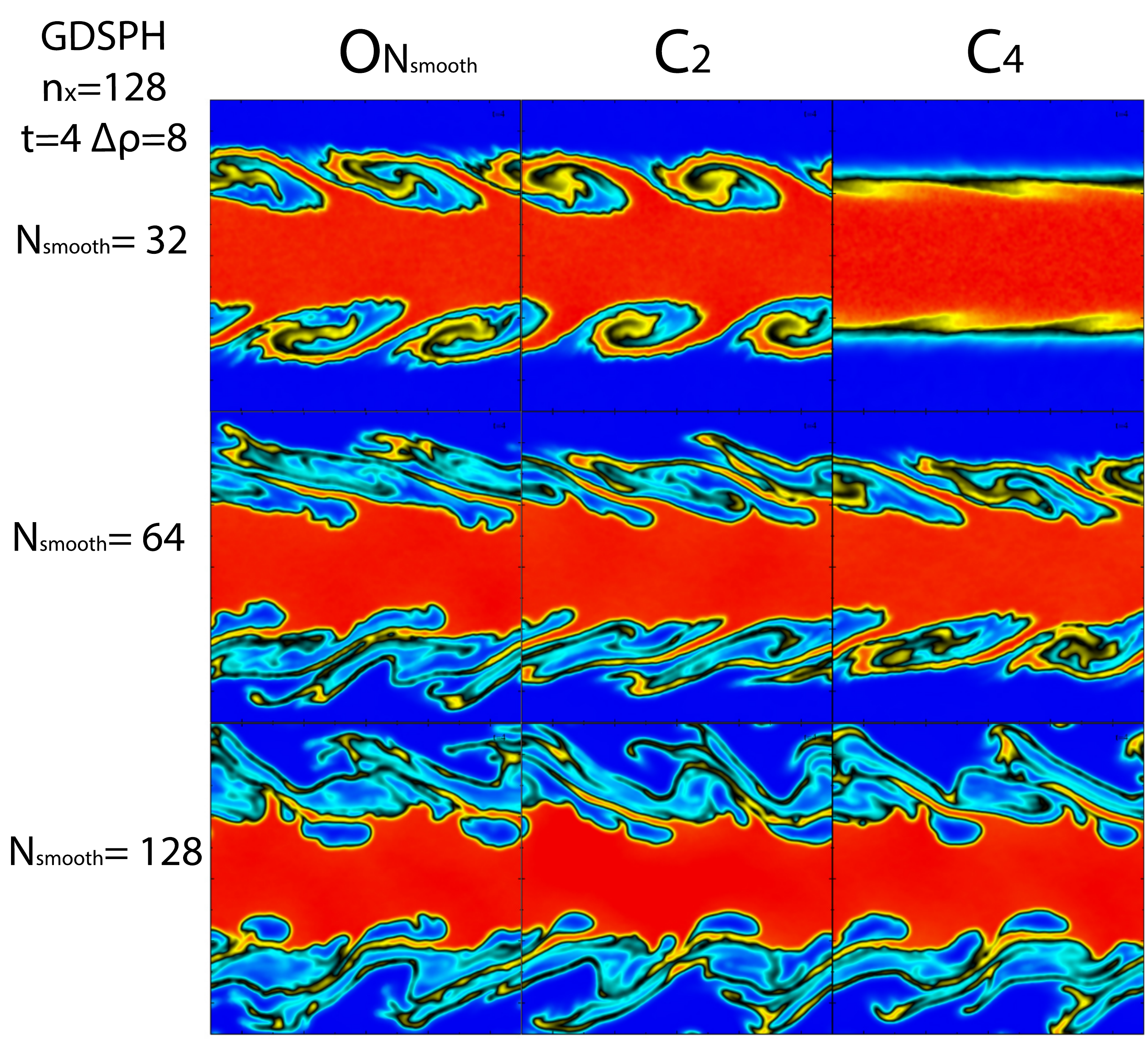}
    \caption{Density rendering of the Kelvin-Helmholtz instability at t=2 (left column) and t=4 (right column) for an initial density contrast of $\Delta \rho=2$ (top figures) and $\Delta \rho=8$ (bottom figures) with the GDSPH method. Here we can see a comparison in the development of the instability between the optimized kernel at a given $N_{smooth}$ and the popular Wendland $C_2$ and $C_4$ kernels. We can see a clear improvement in the optimized kernel, reproducing a similar behavior of the $C_4$ kernel at double the neighbour count ($N_{smooth}$). For the high density contrast $\Delta \rho=8$, there are only minor differences between the different kernels at $N_{smooth}=128$. These can be compared to the higher resolution cases given in Appendix A.}
    \label{fig:khGDSPH}
\end{figure*}
\begin{figure}[]
    \centering
    \includegraphics[width=1.0\linewidth]{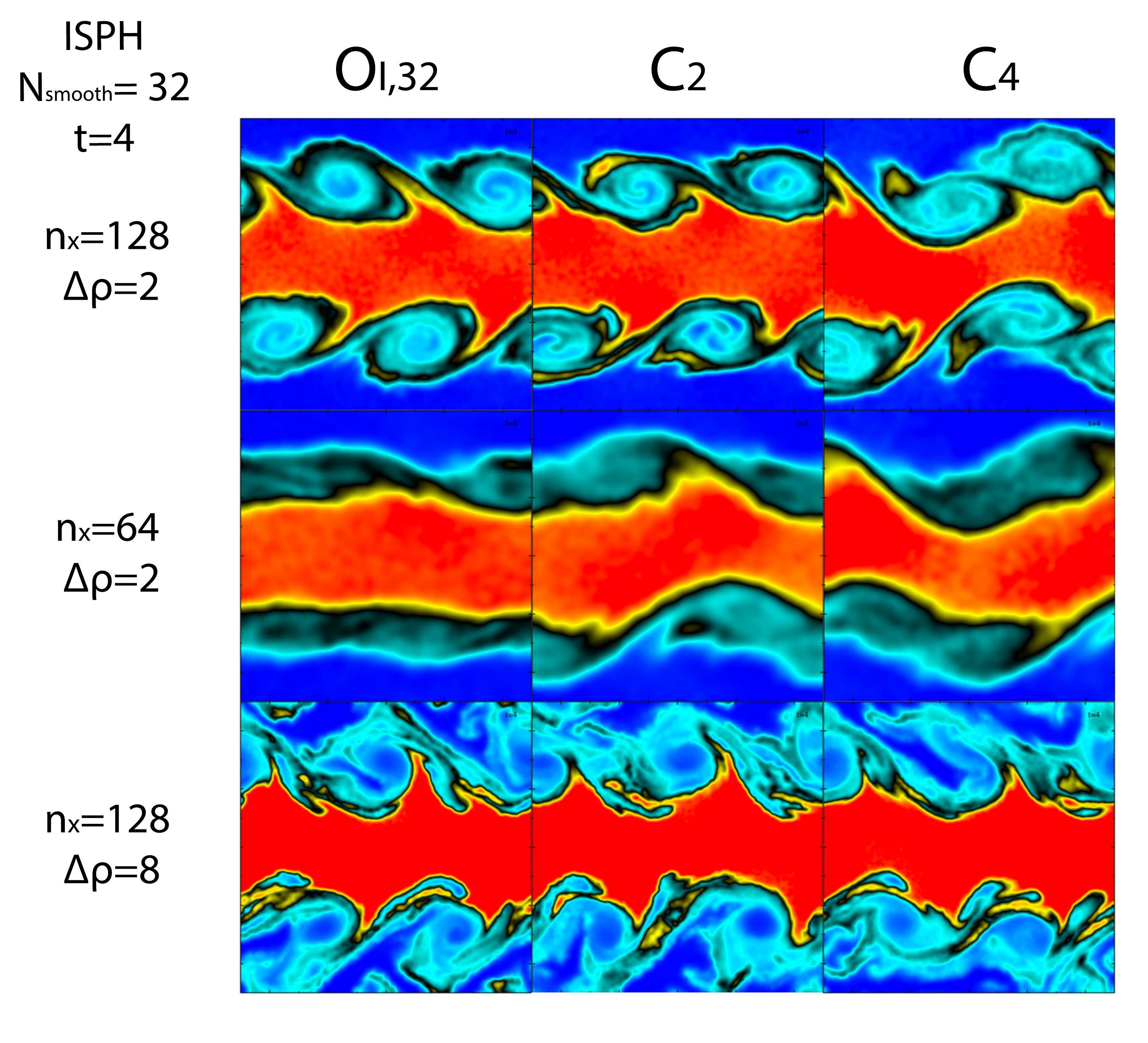}
    \caption{Density rendering of the Kelvin-Helmholtz instability at $t=4$ for $N_{smooth}=32$ with the ISPH method.  Here we can see a comparison in the development of the instability between the optimized kernel ($O{I,32}$) and the popular Wendland $C_2$ and $C_4$ kernels. We have selected three cases where discernible difference can be seen($n_x=128, \Delta \rho = 2 $, $n_x=64, \Delta \rho = 2 $, $n_x=128, \Delta \rho = 8 $). For the $\Delta \rho = 2$ cases we can see that the $C_2$ and $C_4$ kernels exhibit more angular momentum errors (inherent to ISPH) compared to the $O_{I,32}$ kernel, where the inner stream becomes significantly twisted, deviating from the high resolution and high $N_{smooth}$ results (see Appendix A). For  the high density case $n_x=128, \Delta \rho = 8 $, we can see that $C_4$ is unable to generate the same vortex structure in the upper plane of the stream and exhibit some asymmetry around the plane. }
    \label{fig:khISPH}
\end{figure}
The Kelvin-Helmholtz (KH) instability occurs when there is a shearing flow between two fluid interfaces. The instability quickly generates a series of rolling waves at the interface between the two fluids, which eventually breaks down into turbulence, mixing the two fluids. The instability plays a significant role in all kinds of fluids and is a crucial component for any hydrodynamical code to model, as it effectively allow two fluids to mix and generate turbulence. However, traditional SPH methods have been known to struggle with triggering the KH instability, particularly when there are significant density contrasts between the fluids. In this paper we model the KH instability using a similar setup to the one outlined in \cite{2012ApJS..201...18M}, using a thin periodic box ($L=[1,1,2\frac{\sqrt{6}}{n_x}]$), here $n_x$ is the average number of particles in the $x$ direction, and the z boundary is set to be roughly 24 particle spacings. A smoothing function is applied to the density and the shearing velocity accordingly:
\begin{equation}
    f_R=0.5*\left( \tanh{\frac{y+y_{inner}}{\delta}}-\tanh{\frac{y-y_{inner}}{\delta}} \right)
\end{equation}
\begin{equation}
\rho=\rho_{outer}+f_R(\rho_{inner}-\rho_{outer})
\end{equation}
\begin{equation}
v_x=v_{x,outer}+f_R(v_{x,inner}-v_{x,outer})
\end{equation}
Here we set $\rho_{outer}=1$, $\delta=0.025$, $v_{x,outer}=-0.5$ and $v_{x,inner}=0.5$. We simulate the KH instability with two different density contrasts ($\rho_{inner}=\Delta \rho = 2$ and $\rho_{inner}=\Delta \rho = 8$). The pressure is initially uniform ($P_0=\frac{1}{\Gamma \mathcal{M}^2}$), where $\Gamma$ is the adiabatic index and $\mathcal{M}$ is the mach number. We use an adiabatic equation of state with adiabatic index of $\Gamma = 5/3$ and a mach number of $\mathcal{M}=\sqrt{6}/5$. The IC is setup using the IC glass generator as described in \sect \ref{sec:ICgen} for all the different kernels. We run the simulation until $t=4$ for each of the optimized kernels at their specific $N_{smooth}=32,64,128,256$ and SPH method. We then compare these kernels with the Wendland $C_2$ and $C_4$ kernel.
\\ \\
In \Fig \ref{fig:khGDSPH} we show the rendered structure of the KH instability for the GDSPH $\Delta \rho = 2$ and $\Delta \rho = 8$ at $t=2$ and $t=4$. Due to the $N_{smooth}=256$ simulations producing near identical results to the $N_{smooth}=128$ we only show the results of $N_{smooth}=128$. For both the low and high density contrast cases, we can see a clear improvement in the growth of the instability for the optimized kernels, producing similar structure as the $C_4$ kernel at half the $N_{smooth}$.  
\\ \\
In general, the difference between the kernels in the ISPH method is very minor for the KH case, across the different $N_{smooth}$, density contrasts and times. The only exception is for $N_{smooth}=32$ at $t=4$ (see \Fig~\ref{fig:khISPH}), where in the $\Delta \rho = 2$ case the $C_4$ experiences significant non-conservation of angular momentum, leading to the flow rotating (still capturing the rolling vortex waves). For $\Delta \rho = 8$, $t=4$ the optimized kernel, produces a vortex structure (seen at higher $N_{smooth}$ for all kernels), while the $C_4$ kernel fails to capture the vortex. The optimized kernel also produces sharper density structure compared to $C_2$ kernel.
\\ \\
We have also performed more comparisons and a resolution study between GDSPH and ISPH for the KH test, which can be found in Appendix A.
\subsection{Sod shock tube}
\label{sec:sodshock}
\begin{figure}[!h]
    \centering
    \includegraphics[width=1.0\linewidth]{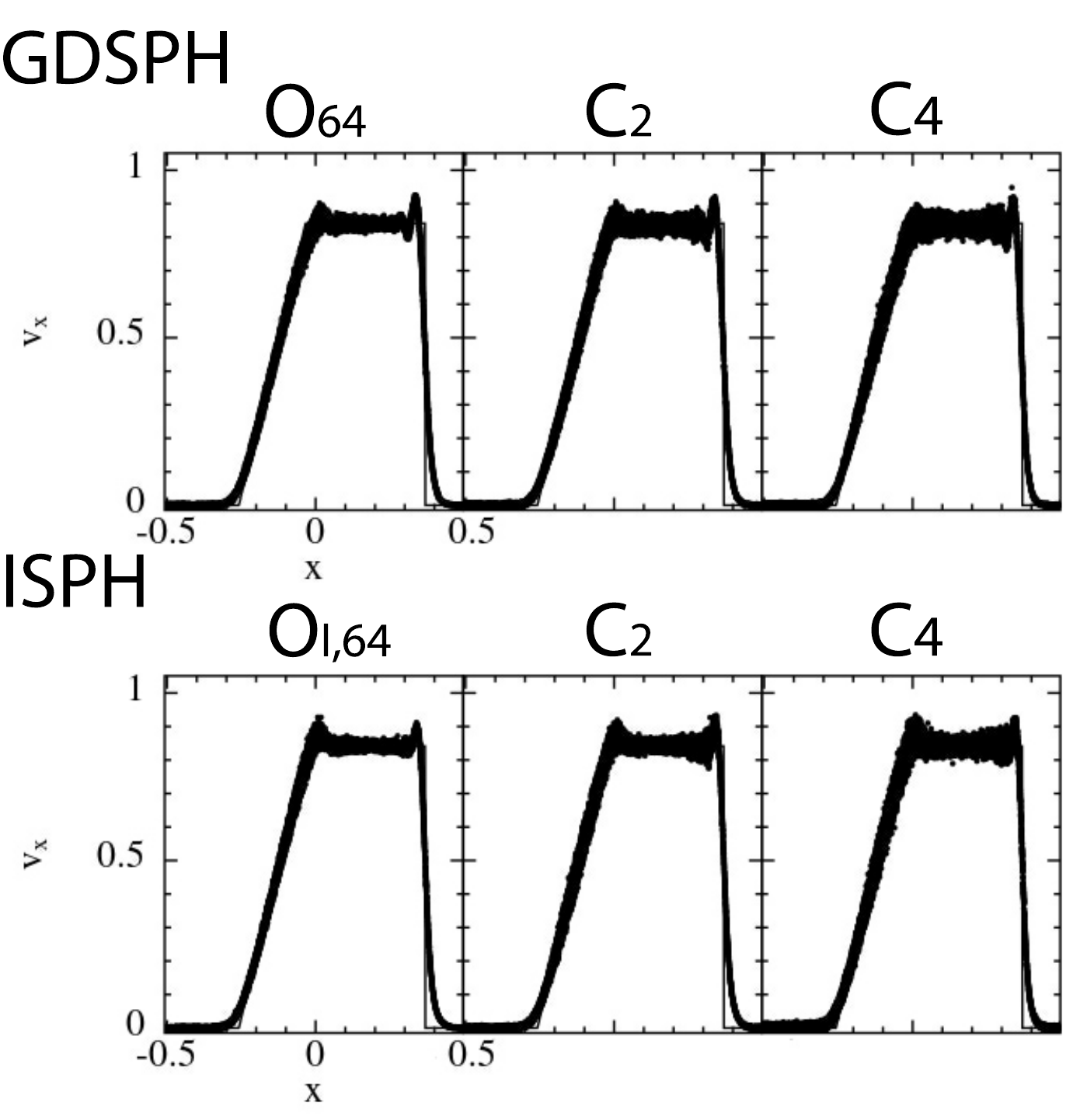}
    \caption{Velocity structure of the Sod-shocktube test at $t=0.2$, performed at $n_x=64$ and $N_{smooth}=64$. The thin black line shows the exact solution. We can see that the optimized kernel exhibit less noise in the post-shock region compared to the C2 and C4 kernel.}
    \label{fig:sodshock}
\end{figure}

Here we test the classic Sod shock tube \citep{1978JCoPh..27....1S}, where a fluid medium is initially divided into two separate regions with different pressures and densities. The initial conditions are $\rho_{left}=1$, $P_{left}=1$ for $x<0$ and $\rho_{right}=0.25$, $P_{right}=0.1795$ for $x>0$. We use an adiabatic equation of state with adiabatic index of $\Gamma = 5/3$. At the start of the simulation, we imagine removing a wall between the two regions creating a discontinuous state between the fluid's generating a shockwave. The produced shockwave has three distinct regions, the right-moving shock front, the left-moving rarefaction wave, and the contact discontinuity. Here there are once again many ways to setup the initial distribution of particles. This case is particularly difficult due to the discontinuous density profile, as it is not possible to follow a discontinuous solution with the smooth density estimate. That is, even if two different density glasses or lattices are set up, the density should still be smoothed at the interface. One could either try one's best to relax a glass following the discontinuous solution or smooth the initial shock by a tiny amount (proportional to the smoothing length). An argument against the complete discontinuous solution (in density) would be that it would rarely happen in a dynamic SPH simulation, as it would always be accompanied with some smoothing. Other quantities such as velocities, thermal energy, etc. of course occur discontinuous in simulations. We choose to smooth the discontinuity on the scale of the smoothing length.
We use a thin periodic 3D box to approximate the 1D test, with $L_x=2.0$ and $L_y=L_z$ is set to be roughly $16$ particle spacings in the low density region. We use $L_x=2.0$ due to using periodic boundary conditions, such that the shock generated at the periodic boundary does not interact with the central shock. We use a resolution of around $n_x=64 \times 2$ in the initial high density region (the factor of 2 is to account for the box extension). We use a constant $\alpha=1$ and $\beta=2$ for the artificial viscosity.
\\ \\
As the velocity shows the biggest difference, we highlight the velocity structure between the different kernels in \fig \ref{fig:sodshock} for $N_{smooth}=64$. From this figure we can see that the optimized kernel's exhibit less noise in the post-shock region compared to the Wendland C2 and C4 kernels for both GDSPH and ISPH. Other than that, there is not much difference between the kernels in this case.
\section{Discussion}
\label{sec:discussion}
In this paper we have generated a set of new optimized kernels through linear-combination of other positive-definite kernels, which coefficients were optimized for the GDSPH method for the hydrostatic glass, the Gresho-Chan vortex and for the ISPH method for the Gresho-Chan vortex. We found that the kernels optimized for the hydrostatic glass perform significantly worse for dynamical cases. We found that all kernels that had regular large oscillations in their Fourier transform performed worse for the Gresho-Chan vortex. The kernels produced by optimizing to the Gresho-Chan vortex all produce quite steady decreasing Fourier transform, with very slight oscillatory behavior, and with a very similar functional form to that of the $C_2$ kernel. However, compared to the $C_2$ kernel, the majority of optimized kernels Fourier transform decreases more rapidly with wave numbers (past $k>3 \pi$). We can see that in general the Fourier transform decreases faster past $k>3 \pi$ for the kernels optimized at higher neighbour number, and decreases faster before $k<3 \pi$ for the kernels optimized at lower neighbour number. Lower power at high wave numbers seem to be important for higher neighbour numbers while a lower power at low wave numbers is important for smaller neighbour numbers. The new optimized kernels produce significant improvements for the GDSPH method, showing an improvement for the Gresho-Chan vortex ($n_x=64$ $t=1$) of $L_{1,error}$ to be $-20\%,-41\%,-31\%,-17\%$ relative to the $L_{1,error}$ of the $C_2$ kernel for $N_{smooth}=32, 64, 128, 256$ respectively. The optimized kernel for $N_{smooth}=64$ gives roughly the same $L_{1,error}$ as that of $N_{smooth}=128$ for the $C_2$ kernel. At the later time, we can see that the $O_{256}$ kernel performs better than $O_{128}$ for all $n_x$ and even better than $O_{64}$ at higher resolution ($n_x=128, 256$), as $O_{256}$ performs quite well even at $t=1$ for both $N_{smooth}=64$ and $N_{smooth}=128$, one could use $O_{256}$ for all $N_{smooth}>64$. The $C_4$ kernel has the steepest improvement for $N_{smooth}=256$, remaining near linear with resolution, while $O_{256}$ starts to flatten out beyond $n_x=256$. This is likely due to the fact that $O_{256}$ reduces zero and first order errors a lot, while $C_4$ reduces higher-order errors more than $O_{256}$. For ISPH we also see improvements of the $L_{1,error}$ in the Greso-Chan vortex, which are $-30\%,-21\%,-11\%,-8\%$ relative to the $L_{1,error}$ of the $C_2$ kernel for $N_{smooth}=32, 64, 128, 256$ respectively. This is visible mainly in a reduction of noise. At the later time, there is much less difference between the kernels at higher $N_{smooth}=128,256$, while a significant improvement can still be seen for $N_{smooth}=32,64$. The $C_2$ kernel is in general the best performing smoothing kernel from the non-optimized kernels for most $N_{smooth}$. At later times at $t=3$, we can see that $C_4$ performs better at $N_{smooth}=64$ for GDSPH, the reason for which is not fully clear and $N_{smooth}=256$ for the ISPH method, where $C_4$ is the best performing kernel of all at $t=3$, likely due to the leading orders being much lower, giving more importance to higher-order errors. The noise of $N_{smooth}=256$ for the ISPH method for $C_4$ is still greater than the $O_{I,256}$ kernel, as can be seen in \Fig \ref{fig:greshoazi}. But the peak of the vortex is slightly higher with the $C_4$ kernel. For the Kelvin-Helmholtz instability the biggest improvement could be seen for the GDSPH method, where the new optimized kernels captured similar results to that of the $C_4$ kernel at double $N_{smooth}$ (both at $\Delta \rho = 2$ and $\Delta \rho = 8$ density contrast). For the ISPH method, the main improvement could be seen in angular momentum conservation of the low $N_{smooth}$ runs, where $O_{I,32}$ performs better than $C_2$ and $C_4$. For the Sod-shocktube test the main improvement of the optimized kernels could be seen as a reduction of the noise in the post-shock region. 
\\ \\
In conclusion, we find that  
\begin{itemize}
\item The linear-combination of kernels generates optimized kernels that significantly improve results, particularly in the low $N_{smooth}$ regime. This can give similar result to doubling the $N_{smooth}$ of previous smoothing kernels ($C_2, C_4$) without any additional computational cost.

\item For GDSPH method we see significant improvement on all test cases with the optimized kernels, while for ISPH the main improvement can be seen in the reduction of noise.

\item We have introduced a new IC generator to generate glass distributions of arbitrary density profiles in SPH. The IC generator, optimized kernels and the test cases in this paper is available at \url{https://github.com/robertwissing/testsuite}

\item We have introduced self-bias corrections to the kernel, that have been adjusted to the glass distribution instead of the lattice distribution of earlier work. This ensures more accurate density interpolation for the glass distribution of positive-definite kernels used in this paper. 

\end{itemize}


\section*{Acknowledgements}
The simulations were performed using the resources from the National Infrastructure for High Performance Computing and Data Storage in Norway, UNINETT Sigma2, allocated to Project NN9477K.

\bibliographystyle{aa}
\bibliography{references}

\begin{appendix}
\section{Differences between GDSPH and ISPH}
\begin{figure*}[]
    \centering
    \includegraphics[width=1.0\linewidth]{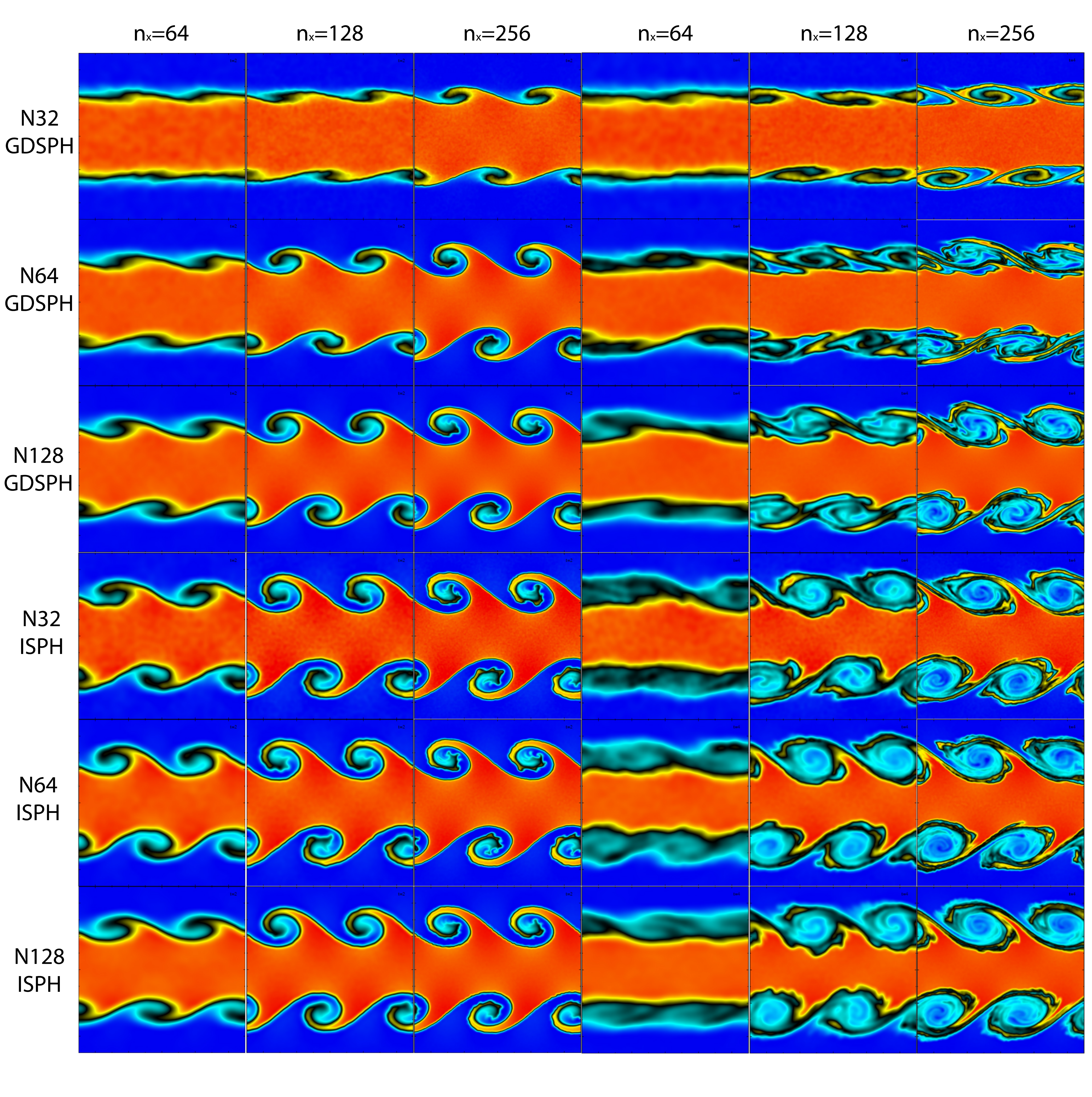}
    \caption{Density rendering of the Kelvin-Helmholtz instability ($\Delta\rho=2$) at $t=2$ (three left columns) and $t=4$ (three right columns) for varying resolution ($n_x=64,128,256$), $N_{smooth}=32,64,128$, and method (GDSPH upper three rows and ISPH bottom three rows). These simulations were all run with the optimized kernel for each $O_{N_{smooth}}$ GDSPH and $O_{I,N_{smooth}}$ for ISPH. We can see that ISPH manages to generate the same vortex structure, regardless of the number of neighbours if the resolution is high enough, while GDSPH struggles with lower number of neighbours even at the highest resolution.}
    \label{fig:APP1}
\end{figure*}

\begin{figure*}[]
    \centering
    \includegraphics[width=1.0\linewidth]{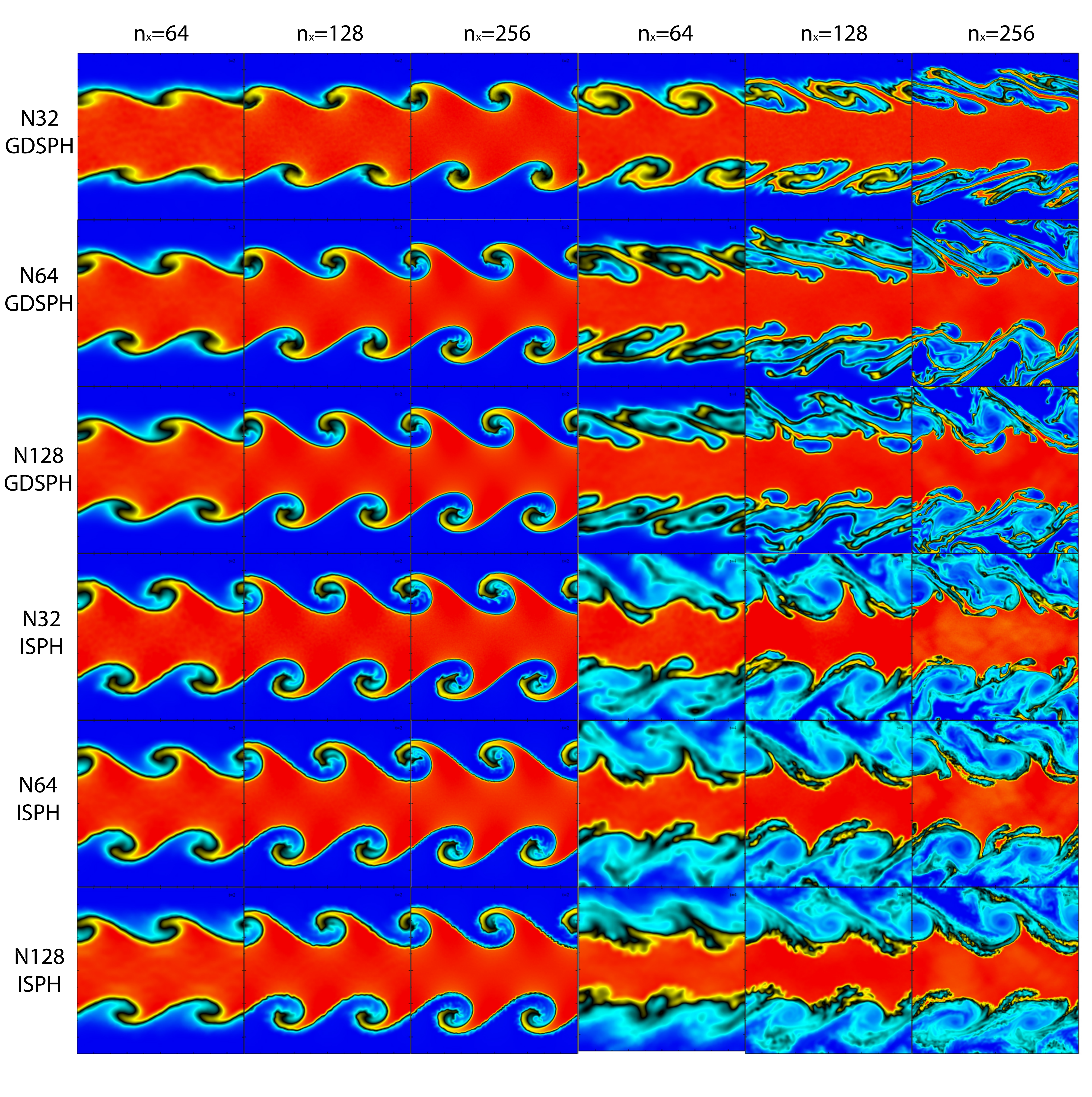}
    \caption{Density rendering of the Kelvin-Helmholtz instability ($\Delta\rho=8$) at $t=2$ (three left columns) and $t=4$ (three right columns) for varying resolution ($n_x=64,128,256$), $N_{smooth}=32,64,128$, and method (GDSPH upper three rows and ISPH bottom three rows). These simulations were all run with the optimized kernel for each $O_{N_{smooth}}$ GDSPH and $O_{I,N_{smooth}}$ for ISPH. Similar as the low density version of \Fig \ref{fig:APP1}, we can see that ISPH captures instability well regardless of neighbour number. We can see additional diffusion from the ISPH runs compared to the GDSPH runs, even with the highest resolution. There also seem to be additional noise in the small-scale density structures as we increase the number of neighbours for ISPH. Potentially related to the issue found in \Fig \ref{fig:khparticlediff}}
    \label{fig:APP2}
\end{figure*}

\begin{figure*}[]
    \centering
    \includegraphics[width=1.0\linewidth]{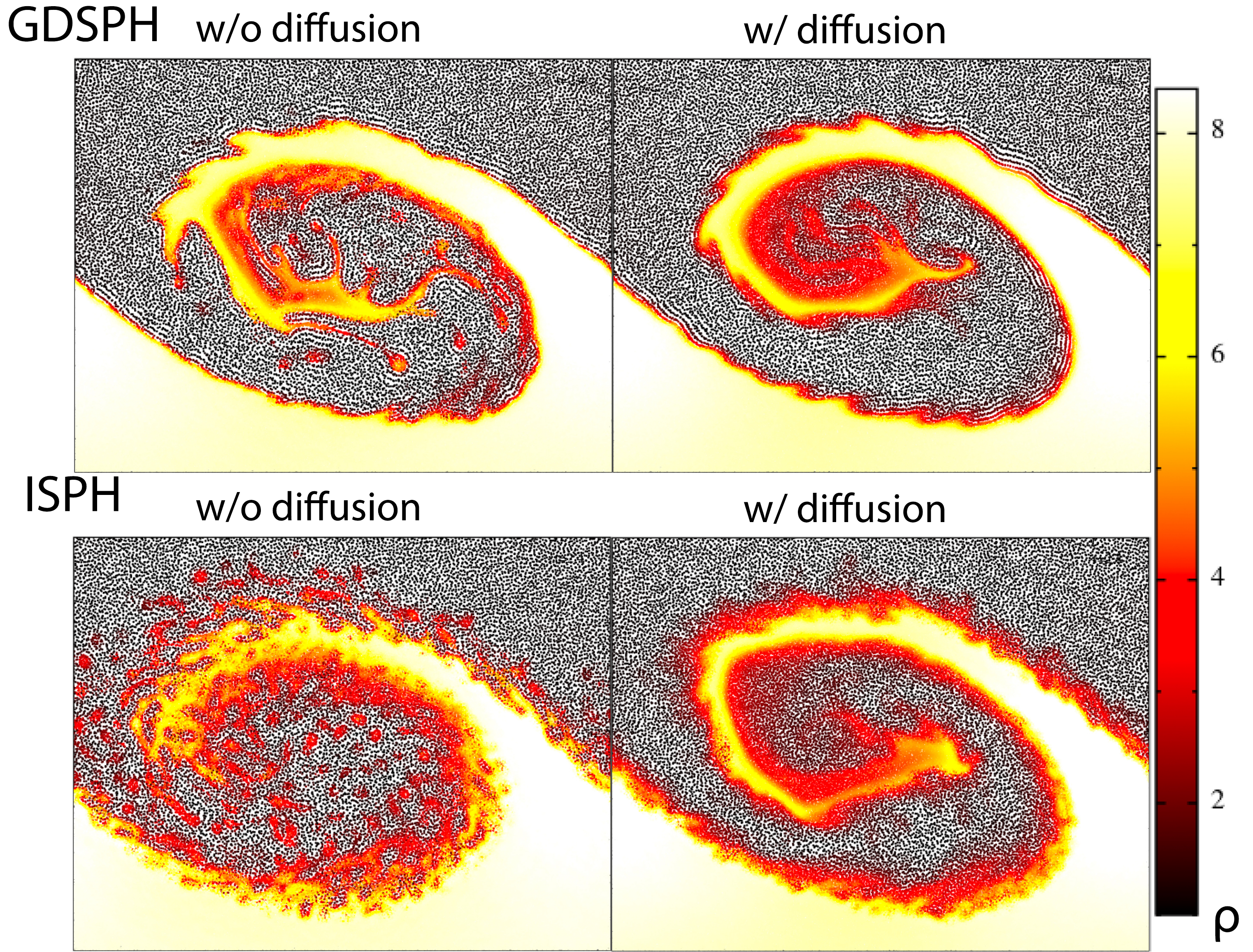}
    \caption{Particle density rendering of Kelvin-Helmholtz instability swirl at $t=2$ for $\Delta \rho = 8$. The figures shows the different behavior of not using diffusion vs.\ using diffusion for the GDSPH and ISPH methods. A significant number of entropy clumps can be seen being generated in the low density medium for ISPH, as the removal of linear gradient together with sharp boundaries causes chaotic noise at the boundary. Lagrangian SPH formalisms that are derived directly from the density estimate, provide full spatial conservation of entropy. However, deviation from the Lagrangian formalism, while still using the traditional density estimate means that we will introduce entropy errors in our solution. GDSPH corrects for first-order errors in entropy making it second-order in entropy. ISPH removes linear-gradient errors and will thus only be first-order accurate with entropy. The issue might be further aggrevated by the use of average gradient kernels within ISPH. Adding thermal diffusion allows for local mixing between the cold and hot phase, and we can see that we get good behavior in both ISPH and GDSPH. Slight numerical surface tension effect can be seen for GDSPH, while none can be seen for ISPH.}
    \label{fig:khparticlediff}
\end{figure*}

In this appendix section we include a resolution study between GDSPH and ISPH for the KH instability using the new optimized kernels for each respective $N_{smooth}$, which can be seen in \Fig \ref{fig:APP1} for $\Delta \rho = 2$ and \Fig \ref{fig:APP2} for $\Delta \rho = 8$. From these figures it is clear that the result for ISPH is highly independent of $N_{smooth}$, producing the same solution for $N_{smooth}=32$ and $N_{smooth}=128$. The GDSPH results are on the other hand, much more dependent on $N_{smooth}$, and exhibit severe damping of the instability at $N_{smooth}=32$. At $N_{smooth}=128$, the GDSPH method produces similar structures to that of the ISPH method at $\Delta \rho = 2$. For high density contrast $\Delta \rho = 8$, we can see that high $N_{smooth}$ for the ISPH method generates more noise at the boundary layer at $t=2$ compared to GDSPH in this case. We can also see that at later time, the high density structures are more dissipate in the ISPH method compared to the GDSPH method. We still see much more consistent behavior in the ISPH method compared to GDSPH method. 
\\ \\
The high density contrast KH instability, is interesting as it is usually accompanied with strong surface tension effects in the Lagrangian formalism of SPH, due to high $E_0$ and $E_1$ errors at the discontinuous boundary. In GDSPH, these errors are reduced by doing a geometric density weighting in the gradient operator, reducing the errors at the density contrast. This sort of correction, deviates from the density estimate derivation of Lagrangian SPH and will thus introduce an entropy error. In GDSPH this is corrected for, such that the method remain second order accurate in entropy. ISPH removes the linear gradient error from it's gradients, which effectively removes more of the "partition" force that exist in SPH and only have $E_0$ error forces to remain ordered. This sort of correction also introduces entropy errors, which would potentially be first-order in this case. In \Fig~\ref{fig:khparticlediff}, we show how one of the KH swirls develops when not using diffusion vs. using diffusion for both the GDSPH and ISPH methods. Significant amounts of entropy bubbles can be seen in the ISPH version without diffusion, and a few can be seen in GDSPH. Adding
thermal diffusion \citep{2010MNRAS.407.1581S} allows for local mixing between the cold and hot phases, and we can see that we get good behavior in both ISPH and GDSPH. Slight numerical surface tension effects can be seen for GDSPH, while none can be seen for ISPH.

\section{Kernel self-bias correction coefficients}
\label{sec:selfbias}
In this section we present the self-correction bias coefficients present in: 
\begin{equation}
    \rho_{a}=\rho_{a,estimate}-\epsilon m_a W(0,h_a)
    \label{eq:biasdens}
\end{equation}
\begin{equation}
    \epsilon(N_{smooth}) =\frac{ c_{i+1} - c_i}{N_{min,i+1}-N_{min,i}}(N_{smooth}-N_{min,i}) + c_i ,
    \label{eq:biascoeff}
\end{equation}
$$    N_{\min,i} \le N_\mathrm{smooth} < N_{\min,i+1},$$
where $N_{min,i}  \in \{16,24,32,48,64,96,128,256,512\}$.
The $c_i$ coefficients can be found in Table \ref{tab:biascoeff} for all kernels.
\begin{table*}[]
\centering
\small
\setlength{\tabcolsep}{3pt} 
\begin{tabular}{|c|c|c|c|c|c|c|c|c|c|c|}
\hline
 & $c_{1}$ & $c_{2}$ & $c_{3}$ & $c_{4}$ & $c_{5}$ & $c_{6}$ & $c_{7}$ & $c_{8}$ & $c_{9}$  \\
\hline
$O_{32}$  & 0.7765625 & 0.85302734 & 0.86884766 & 0.89765625 & 0.92294922 & 0.93652344 & 0.94716797 & 0.96826172 & 0.98261719  \\
\hline
$O_{64}$  & 0.83125305 & 0.93046875 & 0.93710937 & 0.94462891 & 0.95146484 & 0.96054687 & 0.96875 & 0.9828125 & 0.99072266 \\
\hline
$O_{128}$  & 0.83544617 & 0.95097656 & 0.95742187 & 0.96210937 & 0.96582031 & 0.97236328 & 0.97832031 & 0.98867188 & 0.99404297 \\
\hline
$O_{256}$  & 0.84145508 & 0.95 & 0.97675781 & 0.98193359 & 0.98554688 & 0.99023438 & 0.99355469 & 0.99814453 & 0.99931641 \\
\hline
$O_{I,32}$  & 0.67424316 & 0.85517578 & 0.88662109 & 0.88674219 & 0.92255859 & 0.95058594 & 0.97558594 & 1.0 & 1.0 \\
\hline
$O_{I,64}$  & 0.74882812 & 0.85976562 & 0.87744141 & 0.90087891 & 0.91757812 & 0.93232422 & 0.94550781 & 0.96835938 & 0.9828125  \\
\hline
$O_{I,128}$  & 0.80136719 & 0.90957031 & 0.92109375 & 0.93486328 & 0.94765625 & 0.96025391 & 0.97089844 & 0.98574219 & 0.99296875 \\
\hline
$O_{I,256}$  & 0.81279297 & 0.91591797 & 0.92558594 & 0.93730469 & 0.94824219 & 0.95908203 & 0.96845703 & 0.9828125 & 0.99091797 \\
\hline
$C_{2}$  & 0.80507812 & 0.90039062 & 0.91347656 & 0.93349609 & 0.95693359 & 0.97158203 & 0.97675781 & 0.98847656 & 0.99394531 \\
\hline
$C_{4}$  & 0.70195312 & 0.87412109 & 0.91611328 & 0.95019531 & 0.96748047 & 0.98417969 & 0.99013672 & 0.99658203 & 0.99873047 \\
\hline
$C_{6}$  & 0.54921875 & 0.76171875 & 0.83505859 & 0.92001953 & 0.96542969 & 0.98515625 & 0.99130859 & 0.99814453 & 0.99941406 \\
\hline
$C_{8}$  & 0.43124695 & 0.63945312 & 0.72568359 & 0.85019531 & 0.93779297 & 0.98095703 & 0.99003906 & 0.9984375 & 0.99960938 \\
\hline
$WU_{2}$  & 0.94551392 & 0.97011719 & 0.98007812 & 0.98652344 & 0.98964844 & 0.99326172 & 0.99482422 & 0.99707031 & 0.99824219 \\
\hline
$WU_{4}$  & 0.78925781 & 0.96679688 & 0.98798828 & 0.99375 & 0.99619141 & 0.99775391 & 0.99863281 & 0.99941406 & 0.99960938 \\
\hline
$CM_{04}$  & 0.8734375 & 0.89990234 & 0.91533203 & 0.94384766 & 0.95839844 & 0.97167969 & 0.97919922 & 0.99277344 & 1.0 \\
\hline
$CM_{05}$  & 0.74023437 & 0.80107422 & 0.82080078 & 0.85361328 & 0.8921875 & 0.91513672 & 0.92734375 & 0.95458984 & 0.97119141 \\
\hline
$BUH$  & 0.7812439 & 0.86464844 & 0.88203125 & 0.90712891 & 0.93808594 & 0.95830078 & 0.96572266 & 0.98271484 & 0.99082031 \\
\hline
\end{tabular}
\caption{Self-bias correction for GDSPH ($c_{N_{smooth}}$ to be used in \Eq \ref{eq:biascoeff} to calculate correction factor for density \Eq \ref{eq:biasdens}. We can see that all kernels converge towards $1$ at higher $N_{smooth}$.}
\label{tab:biascoeff}
\end{table*}

\begin{table*}[]
\centering
\small
\setlength{\tabcolsep}{3pt} 
\begin{tabular}{|c|c|c|c|c|c|c|c|c|c|}
\hline
 & $f_{1}$ & $f_{2}$ & $f_{3}$ & $f_{4}$ & $f_{5}$ & $f_{6}$ & $f_{7}$ & $f_{8}$ & $f_{9}$  \\
\hline
$O_{32}$  & 1.007 & 1.003 & 1.003 & 1.0 & 0.999 & 0.999 & 0.996 & 0.996 & 0.995  \\
\hline
$O_{64}$  & 1.005 & 0.999 & 1.0 & 0.999 & 0.999 & 0.998 & 0.996 & 0.997 & 0.995 \\
\hline
$O_{128}$  & 1.004 & 0.998 & 0.999 & 0.999 & 0.998 & 0.996 & 0.995 & 0.997 & 0.996 \\
\hline
$O_{256}$  & 1.004 & 1.0 & 0.999 & 0.999 & 0.998 & 0.998 & 0.998 & 0.998 & 0.998 \\
\hline
$O_{I,32}$  & 1.019 & 0.979 & 0.992 & 0.970 & 0.886 & 0.803 & 0.714 & 0.341 & 0.088 \\
\hline
$O_{I,64}$  & 1.007 & 1.001 & 1.002 & 1.001 & 1.0 & 0.999 & 0.996 & 0.997 & 0.996  \\
\hline
$O_{I,128}$  & 1.005 & 1.0 & 1.0 & 1.0 & 1.0 & 1.041 & 0.996 & 0.997 & 0.996 \\
\hline
$O_{I,256}$  & 1.005 & 1.0 & 1.0 & 1.0 & 1.0 & 0.999 & 0.997 & 0.998 & 0.996 \\
\hline
$C_{2}$  & 1.005 & 0.996 & 0.995 & 0.994 & 0.988 & 0.998 & 0.998 & 0.996 & 0.997 \\
\hline
$C_{4}$  & 0.963 & 0.997 & 0.997 & 0.999 & 0.999 & 0.998 & 1.0 & 0.999 & 0.999 \\
\hline
$C_{6}$  & 0.973 & 0.997 & 0.998 & 1.0 & 1.0 & 1.0 & 1.0 & 1.0 & 1.0 \\
\hline
$C_{8}$  & 0.974 & 0.998 & 0.999 & 0.999 & 1.0 & 1.0 & 1.0 & 1.0 & 1.0 \\
\hline
$WU_{2}$  & 0.978 & 0.990 & 0.987 & 0.956 & 0.937 & 0.895 & 0.853 & 0.693 & 0.363 \\
\hline
$WU_{4}$  & 1.003 & 0.998 & 0.998 & 0.997 & 0.997 & 0.993 & 0.994 & 0.995 & 0.993 \\
\hline
$CM_{04}$  & 0.981 & 0.975 & 0.910 & 0.806 & 0.719 & 0.557 & 0.395 & 0.1 & x \\
\hline
$CM_{05}$  & 0.976 & 1.0 & 1.0 & 1.0 & 0.998 & 0.999 & 0.997 & 0.998 & 0.997 \\
\hline
$BUH$  & 0.970 & 0.996 & 0.997 & 0.998 & 0.994 & 0.999 & 0.998 & 0.998 & 0.998  \\
\hline
\end{tabular}
\caption{Multipliers to calculate self-bias correction for ISPH from the values of Table B.1. Correction to self-bias for ISPH is given by using $(f_{i}\cdot c_{i})$ instead of $c_{i}$ in \Eq \ref{eq:biascoeff}. In general the kernels with ISPH have very similar self-bias coefficients to that of GDSPH, but we can see that some kernels in ISPH become increasingly unstable as we increase $N_{smooth}$, like $CM_{04}$, $WU_2$ and $O_{I,32}$, and diverges from the others, which otherwise converges to roughly $1$.}
\label{tab:biascoeff2}
\end{table*}

\section{Kernel tables}
The linear-combination of kernels after optimization are fitted are fitted to a polynomial in the form:
\begin{align}
    P(x) = \sum_{i=0}^{8} c_i x^i\\
    W(q) = f_{norm}\left(1+q^2P(q)\right)
\end{align}
The polynomial coefficients for all the optimized kernels are given in \ref{tab:glasspolycoeff} and  \ref{tab:greshopolycoeff}.
\begin{table*}[h!]
\centering
\small
\setlength{\tabcolsep}{3pt} 
\begin{tabular}{|c|c|c|c|c|c|c|c|c|c|c|}
\hline
 & \text{Norm} & $\alpha_0$ & $\alpha_1$ & $\alpha_2$ & $\alpha_3$ & $\alpha_4$ & $\alpha_5$ & $\alpha_6$ & $\alpha_7$ & $\alpha_8$  \\
\hline
$O_{glass,32}$  & 3.0336 & -12.7824 & 56.1920 & -182.1975 & 398.3558 & -530.5896 & 413.1597 & -173.6614 & 30.5234 & 0.0  \\
\hline
$O_{glass,64}$  & 3.0599 & -7.1464 & 8.4866 & -4.6008 & 22.5183 & -44.3100 & 26.6288 & 6.6825 & -13.2333 & 3.9743 \\
\hline
$O_{glass,128}$  & 2.9693 & -7.104 & 6.2483 & 12.4432 & -28.5090 & 29.4627 & -22.3099 & 11.2888 & -2.5203 & 0.0 \\
\hline
$O_{glass,256}$  & 3.0302 & -7.5780 & 10.7478 & -12.9728 & 55.6311 & -132.8491 & 162.5874 & -110.9071 & 40.5189 & -6.1782 \\
\hline
\end{tabular}
\caption{The polynomial coefficients for all the optimized kernels based on the hydrostatic glass test case.}
\label{tab:glasspolycoeff}
\end{table*}

\begin{table*}[h!]
\centering
\small
\setlength{\tabcolsep}{3pt} 
\begin{tabular}{|c|c|c|c|c|c|c|c|c|c|c|}
\hline
 & \text{Norm} & $\alpha_0$ & $\alpha_1$ & $\alpha_2$ & $\alpha_3$ & $\alpha_4$ & $\alpha_5$ & $\alpha_6$ & $\alpha_7$ & $\alpha_8$  \\
\hline
$O_{32}$  & 3.0702 & -15.1790 & 73.5187 & -257.6388 & 645.4938 & -1078.5898 & 1171.7511 & -797.9519 & 310.1745 & -52.5786  \\
\hline
$O_{64}$  & 3.1160 & -10.7948 & 34.9240 & -85.6680 & 148.6523 & -124.7885 & -11.0621 & 104.2637 & -73.5505 & 17.0239 \\
\hline
$O_{128}$  & 3.2790 & -9.2841 & 21.1601 & -34.5440 & 40.3544 & 24.1265 & -149.8027 & 189.8576 & -105.1688 & 22.3011 \\
\hline
$O_{256}$  & 3.3158 & -6.9471 & 0.6117 & 40.6052 & -109.3938 & 199.6857 & -269.1855 & 230.9765 & -108.5062 & 21.1535 \\
\hline
$O_{I,32}$  & 3.4003 & -37.1051 & 286.5916 & -1109.5005 & 2409.0689 & -3009.6819 & 2067.8803 & -618.2116 & -36.1322 & 46.0904 \\
\hline
$O_{I,64}$  & 3.4319 & -15.2178 & 66.9679 & -205.6619 & 448.2313 & -637.1402 & 569.1468 & -307.2116 & 91.3253 & -11.4399  \\
\hline
$O_{I,128}$  & 3.2796 & -10.3851 & 23.4816 & -17.8825 & -44.4497 & 195.2885 & -338.6360 & 309.5800 & -146.1530 & 28.1562 \\
\hline
$O_{I,256}$  & 3.1813 & -10.9758 & 34.0224 & -79.6218 & 145.2028 & -156.0396 & 69.5784 & 17.6550 & -28.5200 & 7.6987 \\
\hline
\end{tabular}
\caption{The polynomial coefficients for all the optimized kernels based on the Gresho-Chan test case.}
\label{tab:greshopolycoeff}
\end{table*}

\begin{table*}[h!]
\centering
\begin{tabular}{|c|c|c|c|c|c|c|c|c|c|c|c|c|c|c|}
\hline
 & $O_{g,64}$ & $O_{g,128}$ & $O_{g,256}$ & $CM_{04}$ & $WU_2$ & $BUH$ & $CM_{05}$ & C2 & C4 & C6 & C8 & $WU_4$ \\
\hline
$O_{32}$  & -1.630 & 0.604 & 0.291 & 0.223 & 0.030 & 0.138 & 0.289 & -0.173 & -0.080 & 0.046 & -0.003 & 0.016 \\
\hline
$O_{64}$  & -0.032 & 0.155 & 0.000 & 0.000 & -0.039 & 0.045 & -0.140 & 0.155 & 0.097 & -0.017 & -0.003 & -0.047 \\
\hline
$O_{128}$  & 0.424 & -0.158 & -0.068 & -0.011 & -0.113 & 0.034 & -0.215 & 0.278 & 0.129 & -0.035 & -0.003 & 0.071 \\
\hline
$O_{256}$  & 0.603 & -0.087 & 0.227 & 0.075 & -0.058 & -0.189 & -0.355 & 0.371 & 0.102 & -0.023 & -0.008 & 0.261 \\
\hline
$O_{I,32}$  & -1.397 & 0.530 & 0.023 & 0.008 & 0.018 & -0.171 & 0.099 & -0.073 & -0.023 & 0.007 & 0.000 & 0.009 \\
\hline
$O_{I,64}$  & -1.397 & 0.530 & -0.088 & -0.005 & -0.014 & 0.018 & 0.047 & -0.027 & 0.094 & -0.089 & 0.034 & -0.027 \\
\hline
$O_{I,128}$  & -0.670 & 0.377 & 0.000 & 0.000 & 0.051 & -0.009 & 0.085 & -0.136 & 0.015 & -0.013 & 0.004 & -0.020 \\
\hline
$O_{I,256}$  & -1.131 & 0.536 & -0.012 & -0.005 & 0.046 & 0.118 & 0.012 & -0.152 & -0.027 & 0.006 & 0.006 & 0.025 \\
\hline
\end{tabular}
\caption{Contribution of each kernel in the linear combination used to optimize all kernels for the Gresho-Chan test case.}
\label{tab:greshokernel}
\end{table*}

\section{Code implementation}
A simple code snippet of the kernel function is included for clarification and for easier implementation, for both GDSPH and ISPH.
\begin{figure*}[t]
\begin{Ccode}
/** GDSPH VERSION
 * @brief kernelOptimized is an optimized kernel for SPH, with different form for different nsmooth
 *
 * This kernel is explained and defined in Wissing (2025).  The kernel
 * is a linear combination of several positive definite kernels,
 * which coeffecients have been optimized to give the best result for the gresho chan vortex
 * r = |dx|/H
 * And for us, H = 2*h_smooth
 *
 * Includes a correction for self-interactions.
 * Which give density of 1 for a glass relaxed with the kernel.
 *
 * NOTE: this kernel should not be called for r > 1
 * @param ar2 = (|dx|/h)^2 = (2r)^2
 * @return (pi h^3) W
 */
inline double kernelOptimized(double ar2, int nSmooth, double mnorm)
{
  double ak;
  double Rkern2=4.0;
  double Wzero=1.0,norm;
  double c0,c1,c2,c3,c4,c5,c6,c7,c8;
  //Wzero values for piecewise function
  double n16,n24,n32,n48,n64,n96,n128,n192,n256,n512;
  if (nSmooth <= 32) //O_I,32 Wissing et al. 2025
    {
      n16=0.68671868118; n24=0.8375001518; n32=0.87988276971;
      if(nSmooth<=16) Wzero=(n16-0.40)/(14)*(nSmooth-2)+0.40;
      if(nSmooth<=24) Wzero=(n24-n16)/8*(nSmooth-16)+n16;
      if(nSmooth<=32) Wzero=(n32-n24)/8*(nSmooth-24)+n24;
      norm=(3.4003319462332073/8./M_1_PI);
      c0=-37.10514490418083;
      c1=286.59163136982374;
      c2=-1109.5005318509457;
      c3=2409.0689549866306;
      c4=-3009.6818640918887;
      c5=2067.880326832294;
      c6=-618.2115694970241;
      c7=-36.132229318136154;
      c8=46.09042647341556;
    }
  else if (nSmooth <= 96) //O_I,64 Wissing et al. 2025
    {
      n32=0.87900413315; n48=0.90136718636; n64=0.91787082742; n96=0.9315429323;
      if(nSmooth<=48) Wzero=(n48-n32)/16*(nSmooth-32)+n32;
      if(nSmooth<=64) Wzero = (n64-n48)/16*(nSmooth-48)+n48;
      if(nSmooth<=96) Wzero = (n96-n64)/32*(nSmooth-64)+n64;
      norm=(3.4318494020523547/8./M_1_PI);
      c0=-15.217780383519472;
      c1=66.96785864668652;
      c2=-205.6618857567194;
      c3=448.23132682213446;
      c4=-637.1401597525037;
      c5=569.1467883955034;
      c6=-307.2115693839176;
      c7=91.32532556233177;
      c8=-11.439904149996446;
    }
  else //O_I,256 Wissing et al. 2025
    {
      n96=0.95820351086; n128=0.96523439238; n192=0.9764745855; n256=0.98056677343; n512=0.98710987224;
      if(nSmooth<=128) Wzero = (n128-n96)/32*(nSmooth-96)+n96;
      if(nSmooth<=192) Wzero = (n192-n128)/64*(nSmooth-128)+n128;
      if(nSmooth<=256) Wzero = (n256-n192)/64*(nSmooth-192)+n192;
      if(nSmooth<=512) Wzero = (n512-n256)/256*(nSmooth-256)+n256;
      norm=(3.181335291127915/8./M_1_PI);
      c0=-10.975827966498654;
      c1=34.022346938397256;
      c2=-79.62182205737668;
      c3=145.2028297298656;
      c4=-156.0395550888878;
      c5=69.57842722118721;
      c6=17.65497978738222;
      c7=-28.52004202463561;
      c8=7.698663460565884;
    }

  if (ar2 <= 0) ak = norm*Wzero;
  else if (ar2 >= Rkern2) ak = 0.0;
  else {
    double au = sqrt(ar2*0.25);
    ak = norm*(1+au*(au*(c0+au*(c1+au*(c2+au*(c3+au*(c4+au*(c5+au*(c6+au*(c7+c8*au))))))))));
//  Gradient adk=(pi h^5/|dx|^2) (dx.dot.gradW)
//  adk = (norm/4.)*((c0*2+au*(c1*3+au*(c2*4+au*(c3*5+au*(c4*6+au*(c5*7+au*(c6*8+au*(c7*9+c8*10*au)))))))));
  }
  return ak;
}
\end{Ccode}
\end{figure*}
\begin{figure*}[t]
\begin{Ccode}
/** ISPH VERSION
 * @brief kernelOptimized is an optimized kernel for SPH, with different form for different nsmooth
 *
 * This kernel is explained and defined in Wissing (2025).  The kernel
 * is a linear combination of several positive definite kernels,
 * which coeffecients have been optimized to give the best result for the Gresho-Chan vortex
 * r = |dx|/H
 * And for us, H = 2*h_smooth
 *
 * Includes a correction for self-interactions.
 * Which give density of 1 for a glass relaxed with the kernel.
 *
 * NOTE: this kernel should not be called for r > 1
 * @param ar2 = (|dx|/h)^2 = (2r)^2
 * @return (pi h^3) W
 */
inline double kernelOptimized(double ar2, int nSmooth, double mnorm)
{
  double ak;
  double Wzero=1.0,norm;
  double Rkern2=4.0;
  double c0,c1,c2,c3,c4,c5,c6,c7,c8;
  //Wzero values for piecewise function
  double n16,n24,n32,n48,n64,n96,n128,n192,n256,n512;
  if (nSmooth <= 32) //O_32 Wissing et al. 2025
    {
      n16=0.7765625; n24=0.85302734; n32=0.86884766;
      if(nSmooth<=16) Wzero=(n16-0.40)/(14)*(nSmooth-2)+0.40;
      if(nSmooth<=24) Wzero=(n24-n16)/8*(nSmooth-16)+n16;
      if(nSmooth<=32) Wzero=(n32-n24)/8*(nSmooth-24)+n24;
      norm=(3.0702399143211334/8./M_1_PI);
      c0=-15.178969768800927;
      c1=73.51866968356427;
      c2=-257.63874599152774;
      c3=645.4937717480914;
      c4=-1078.589778178365;
      c5=1171.7510466338265;
      c6=-797.951908267712;
      c7=310.1744769215709;
      c8=-52.578562780647644;
    }
  else if (nSmooth <= 96) //O_64 Wissing et al. 2025
    {
      n32=0.93710937; n48=0.94462891; n64=0.95146484; n96=0.96054687;
      if(nSmooth<=48) Wzero=(n48-n32)/16*(nSmooth-32)+n32;
      if(nSmooth<=64) Wzero = (n64-n48)/16*(nSmooth-48)+n48;
      if(nSmooth<=96) Wzero = (n96-n64)/32*(nSmooth-64)+n64;
      norm=(3.1160348893350283/8./M_1_PI);
      c0=-10.794751909300937;
      c1=34.923977443679455;
      c2=-85.66799299517906;
      c3=148.65227923861258;
      c4=-124.78847965696686;
      c5=-11.062047962891707;
      c6=104.26365151520089;
      c7=-73.55050556719418;
      c8=17.023869894040175;
    }
      else //O_256 Wissing et al. 2025
    {
      n96=0.99023438; n128=0.99355469; n192=0.99746094; n256=0.99814453; n512=0.99931641;
      if(nSmooth<=128) Wzero = (n128-n96)/32*(nSmooth-96)+n96;
      if(nSmooth<=192) Wzero = (n192-n128)/64*(nSmooth-128)+n128;
      if(nSmooth<=256) Wzero = (n256-n192)/64*(nSmooth-192)+n192;
      if(nSmooth<=512) Wzero = (n512-n256)/256*(nSmooth-256)+n256;
      norm=(3.3158322817560615/8./M_1_PI);
      c0=-6.94709887927258;
      c1=0.6116847826437315;
      c2=40.605173332308;
      c3=-109.39383705074306;
      c4=199.68569324945807;
      c5=-269.18551329586074;
      c6=230.97653909776236;
      c7=-108.50616849023058;
      c8=21.153527253935298;
    }

  if (ar2 <= 0) ak = norm*Wzero;
  else if (ar2 >= Rkern2) ak = 0.0;
  else {
    double au = sqrt(ar2*0.25);
    ak = norm*(1+au*(au*(c0+au*(c1+au*(c2+au*(c3+au*(c4+au*(c5+au*(c6+au*(c7+c8*au))))))))));
//  Gradient adk=(pi h^5/|dx|^2) (dx.dot.gradW)
//  adk = (norm/4.)*((c0*2+au*(c1*3+au*(c2*4+au*(c3*5+au*(c4*6+au*(c5*7+au*(c6*8+au*(c7*9+c8*10*au)))))))));
  }
  return ak;
}
\end{Ccode}
\end{figure*}

\end{appendix}

\end{document}